\documentclass{article}

\usepackage{arxiv}

\usepackage[utf8]{inputenc} 
\usepackage[T1]{fontenc}    
\usepackage{hyperref}       
\usepackage{url}            
\usepackage{booktabs}       
\usepackage{amsfonts}       
\usepackage{nicefrac}       
\usepackage{microtype}      
\usepackage{lipsum}
\usepackage{xr}             
\usepackage{placeins}

\usepackage{longtable}
\usepackage{fancyhdr}
\usepackage{graphicx}
\graphicspath{{media/}{Figures/}}
\usepackage{tabularx}
\usepackage{listings}
\usepackage{fancyvrb}

\usepackage[numbers]{natbib}
\usepackage{amsmath,amssymb}
\usepackage{amsthm}
\usepackage{mathtools}
\usepackage{xspace}
\usepackage[noend]{algorithmic}
\usepackage[ruled,vlined]{algorithm2e}
\usepackage{enumitem}
\usepackage{pdfpages}
\usepackage{chngcntr}
\usepackage{mdframed}
\usepackage{caption}
\usepackage{color}
\usepackage{xcolor}
\usepackage{subfigure}
\usepackage[most]{tcolorbox}
\usepackage{float}
\usepackage{alltt}
\usepackage{soul}
\usepackage{multirow}
\usepackage{csquotes}
\usepackage{upgreek}
\usepackage[normalem]{ulem}

\definecolor{lightred}{rgb}{1,0.9,0.9}

\newcommand{\CVaR}{\ensuremath{\mathrm{CVaR}}\xspace}
\newcommand{\PFCZM}{PF-CZM\xspace}
\newcommand{\epsbar}{\bar{\varepsilon}}

\title{Isotropic Failure Without Symmetry: Stochastic Fracture Topology Optimization Converges on and Rivals Bioinspired Architected Materials}

\author{
  Jaime A. Berkovich\textsuperscript{1,2} \quad
  Markus J. Buehler\textsuperscript{1,3,4,5,\#} \\
  \\
  \textsuperscript{1}Laboratory for Atomistic and Molecular Mechanics (LAMM)\\
  \textsuperscript{2}Department of Materials Science and Engineering\\
  \textsuperscript{3}Department of Civil and Environmental Engineering, \\
  \textsuperscript{4}Department of Mechanical Engineering, \\
  \textsuperscript{5}Center for Computational Science and Engineering, \\
  Schwarzman College of Computing, \\
  Massachusetts Institute of Technology, Cambridge, MA 02139, USA \\ \\
  \textsuperscript{\#}Corresponding author:
  \texttt{mbuehler@MIT.EDU}
}

\begin{document}
\maketitle

\begin{abstract}
Brittle materials fail through interactions between architecture, defects, and loading direction. We introduce CRAX, a JAX-based topology optimization framework differentiating through phase-field fracture simulations with spatially correlated strength and toughness disorder. For periodic materials, optimization over loading directions yields a cell without point symmetry with the lowest maximum directional variation across simulated stiffness, strength, work, and apparent toughness among tested cells. Printed specimens of this cell show no detectable directional dependence in strength or work across 12 orientations. Under uniaxial loading, another cell matches the best canonical lattices in simulated work and lies with them on the work--strength Pareto front; a local-volume constraint improves single-member-loss tolerance at similar work. CRAX discovers foam-like and nacre-like motifs without templates. On seven of eight structural benchmarks, selected CRAX designs absorb 1.31–4.10 times the work of stiffness-optimal counterparts, and the same solver recovers six material and disorder parameters within 5\% from synthetic fracture histories.
\end{abstract}


\section{Introduction}\label{sec:main}

Advances in inverse design for architected materials have increased the breadth of their achievable properties and behaviors~\cite{lee_2023_DataDriven}. For structural metamaterials, a microstructure that absorbs large amounts of energy during deformation without sacrificing much stiffness is desirable, yet this goal runs directly into the classic strength--toughness conflict~\cite{ritchie_2011_Conflicts}: materials that withstand high stresses usually absorb little energy before failing catastrophically, whereas materials that absorb large amounts of energy are usually not strong. The problem is sharpest for brittle and quasi-brittle constituents, such as ceramics and stiff polymers. Brittle parts fail catastrophically when a single flaw reaches critical size, so reliability is set by toughness and by the statistics of its flaws~\cite{weibull_1951_Statistical}. Established toughening levers (inclusions, secondary phases that form ceramic-matrix composites, and changes in chemistry and synthesis~\cite{gavalda-diaz_2025_Toughening}) leave one comparatively unexplored, which printing in complex geometries now makes accessible: architecture, the geometry of the material itself.

A design strategy for tough metamaterials made of brittle constituents therefore requires fracture awareness. Fracture-aware microstructure optimization remains limited, however, in part because fracture is path-dependent and stochastic. It is commonly framed as crack blunting, where the design goal is to prevent or delay the growth of a crack under a prescribed stress or strain, as in~\cite{singh_2026_Designing}. In bulk brittle and quasi-brittle materials, however, the locations where cracks nucleate are not known a priori, since local strength and toughness vary with inhomogeneities such as defects, impurities, and additives; phase-field models can represent this variation as cross-correlated random fields of strength and toughness~\cite{wu_2023_Phasefield}. We therefore argue for a robust fracture-aware design approach accounting for crack \emph{nucleation} under spatially correlated material disorder, not only for the growth of a crack from a known flaw. While the elastic and strength scaling of common cellular architectures is well established~\cite{gibson_1997_Cellular}, their fracture response, which largely controls the energy absorbed during failure, requires separate analysis~\cite{shaikeea_2022_Toughness}. Geometric disorder alone can raise the toughness of a two-dimensional lattice to more than 2.6 times that of the ordered lattice of equal density~\cite{fulco_2025_Disorder}. Architecture also determines how fracture resistance depends on loading direction. In practice that direction is rarely known in advance, and a brittle part is only as strong as its weakest direction. For example, a rule-based quasi-periodic tiling of printed polylactic acid was shown to absorb over two times the fracture energy of a density-matched honeycomb, with peak load and fracture energy varying minimally across three orientations~\cite{gao_2026_TopologyEnriched}. Other approaches impose elastic isotropy through symmetry constraints while optimizing for fracture initiation~\cite{singh_2026_Designing}, or report similar fracture toughness along two orthogonal crack directions for spiral microstructures~\cite{wang_2025_Architected}. Periodic cellular materials have a clear advantage over disordered tilings: a single cell can be analyzed via homogenization.

Biological systems have long since leveraged nanoscale and microscale size effects to their advantage, offering ceramic-based structural materials that resolve the strength--toughness conflict, among them organic--inorganic nanocomposites such as bone and nacre~\cite{wegst_2015_Bioinspired}. Evolution has already pushed otherwise brittle materials into tougher regimes, and evolutionary~\cite{da_2020_Topology} and bioinspired~\cite{wegst_2015_Bioinspired} approaches have accordingly been used to design architected materials. Bioinspired motifs are, however, heuristics, transplanted from an organism into a new constituent, fabrication route, and loading condition, and carry little information about the disorder of the material they are built from or how damage evolves in it.

Outside the realm of human-acquired heuristics, machine learning has enabled the de novo design of composites and architected materials through surrogate and generative models~\cite{gu_2018_De, gu_2018_Bioinspired, lee_2023_DataDriven}. Recently, large language models (LLMs) and autonomous agents have entered materials design, as multi-agent systems that generate and refine hypotheses~\cite{ghafarollahi_2024_SciAgents} and as an agent that builds and validates its own simulator~\cite{buehler_2026_Models}. The reliability of these systems is largely constrained by the physical models they use as truth sources. For the fracture of brittle architected materials those models are typically expensive, deterministic, and non-differentiable, so data-driven and agentic design has rarely targeted fracture under material disorder: learned models have predicted fracture paths under a random toughness field~\cite{quek_2024_Approximating} and optimized them in geometrically disordered lattices~\cite{karapiperis_2023_Prediction}, but none has designed under random strength and toughness together. A differentiable, disorder-aware fracture solver would be useful both as a design method in its own right and as an instrument for future agentic systems.

Topology optimization (TO)---a method that distributes a prescribed budget of material over a design domain to extremize a performance functional subject to governing equations and design constraints~\cite{bendsoe_1989_Optimal}---has strongly embraced differentiable physics in recent years. For the well-studied compliance problem, the system is self-adjoint and gradients are cheap, whereas failure-aware TO is harder by comparison. Stress-based TO was followed by damage-aware and fracture-aware formulations that couple level-set, density-based, or evolutionary design updates to crack extension, nonlocal damage, or phase-field fracture, as reviewed by Yvonnet and Da~\cite{yvonnet_2024_Topology}, including bidirectional evolutionary structural optimization with phase-field fracture in periodic quasi-brittle composites~\cite{da_2020_Topology}; differentiable peridynamic fracture has since been demonstrated~\cite{schmitz_2026_Topology}. Most recently, Singh and Sigmund~\cite{singh_2026_Designing} designed 2D periodic materials with delayed fracture initiation from the near-tip stresses around a single pre-existing crack under K-field loading, subject to compliance symmetry, and Holm et al.~\cite{holm_2026_Mode} maximized the mode-I fracture toughness of periodic beam lattices, finding that, normalized by cell size, classical triangular and Kagome lattices were not surpassed. Most of these methods design structures and the few that design periodic architected cells~\cite{da_2020_Topology, singh_2026_Designing, holm_2026_Mode} assume deterministic material properties and, typically, one crack orientation and length relative to the unit cell.

Compliance TO has long accounted for uncertainty in loads, material stiffness, and manufacturing, and has optimized risk measures such as the conditional value-at-risk under random stiffness fields and load directions~\cite{eigel_2018_Risk}; the linear elastic state problem, one linear solve per realization, keeps such formulations tractable. Fracture, on the other hand, is nonlinear, irreversible, and path-dependent, so these simplifications do not apply, and the spatially correlated variation of strength and toughness that decides where cracks nucleate has not historically been part of the design loop. Uncertainty has only recently entered fracture-aware TO, through interval models of load and performance~\cite{zhao_2025_Interval} and through Monte Carlo ensembles of random defects with averaged or tail-weighted fracture energy~\cite{li_2027_Topology}, all for structures rather than periodic metamaterials. To the best of our knowledge, no existing TO method combines (i) fracture awareness with post-peak softening, (ii) spatially correlated strength and toughness disorder, and (iii) exact sensitivities through a staggered, history-dependent solve.

We ask whether failure that depends little on load direction can be optimized for without imposing symmetry, and what an optimizer builds when given only the fracture physics and the disorder of its constituent. To this end, we introduce CRAX, a GPU-accelerated, \uline{CR}ack-aware, J\uline{AX}-based topology-optimization method that designs periodic microstructures de novo, without a template, for the \emph{distribution} of their fracture response. CRAX builds on density-based solid isotropic material with penalization (SIMP) and couples a length-scale-insensitive phase-field cohesive-zone model (PF-CZM)~\cite{wu_2017_Unified, wu_2018_Length}, a cohesive variant of the variational approach to fracture~\cite{bourdin_2000_Numerical}, automatic differentiation~\cite{bradbury2018jax, xue_2023_JAXFEM, ani_2026_Matrixfree} through batches of staggered, history-dependent fracture solves on lognormal strength and toughness fields with spatial correlation and, where the load direction is unknown, a random load direction for each realization, and a conditional value-at-risk ($\CVaR$) objective~\cite{rockafellar_2000_Optimization, kouri_2016_RiskAverse} that grades each design on the worst realizations (simulation replicates using stochastic property fields) of its absorbed work rather than on its average. We derive exact implicit adjoints for the bound-constrained damage problem and verify them against finite differences; the designs reported here were optimized with straight-through estimators in place of that adjoint, and a re-optimization study with the exact adjoint is reported in the Supplementary Information (SI, Section~S13).

Our main contributions are a design method and three findings obtained with it. Unlike fracture-aware optimizers tied to one
length scale, boundary condition, and objective, CRAX differentiates through batches of phase-field fracture
simulations on correlated random strength and toughness fields, with no prescribed crack, and grades designs by the
conditional value-at-risk of absorbed work; on seven of eight standard problems its designs absorb 1.31 to 4.10 times
the work of their stiffness-optimal twins at equal volume (\S\ref{sec:framework}, \S\ref{sec:benchmarks}). The findings are:
\begin{itemize}[leftmargin=*, itemsep=2pt, topsep=4pt]
  \item \textbf{Isotropic failure without symmetry.} Optimized over random load directions, CRAX finds a
  cell without point symmetry (CRAX-W$\Omega$) with the lowest maximum direction dependence across simulated stiffness,
  strength, work, and apparent toughness among tested cells; when tiled, printed, and mechanically tested in 12 directions, its strength and work showed no detectable
  direction dependence (\S\ref{sec:direction}, \S\ref{sec:printed}).
  \item \textbf{A work-optimized cell on the work--strength Pareto front.} Optimized for absorbed work under uniaxial
  strain, CRAX-WU matches the best bioinspired lattices in absorbed work and lies on the work--strength
  Pareto front of the tested cells, both in silico and experimentally (\S\ref{sec:canonical}, \S\ref{sec:printed}).
  \item \textbf{Motif classes without a template.} From a nearly uniform density, CRAX converges on closed-cell foam
  and nacre-like brickwork (\S\ref{sec:novelty}). The characteristic length sets the scale of the architecture: optimized at $\ell_{ch}/L = 0.10$ instead
  of 1.64, designs have 2--4 instead of 16--23 pores per cell and walls comparable with $\ell_{ch}$
  (SI, Section~S17.2).
\end{itemize}
We also show a trade-off between damage spread and crack isotropy, in which cells that spread damage had the most
direction-dependent toughness (Spearman $\rho = -0.90$, eight designs; \S\ref{sec:mechanisms}), and that the same solver recovers material
properties and, blind, the disorder statistics from fracture histories by gradient descent rather than Bayesian
sampling~\cite{khodadadian_2020_Bayesian}, all six parameters within 5\,\% (\S\ref{sec:inverse}).

\section{Results}\label{sec:results}

\subsection{Disorder-aware fracture design}\label{sec:framework}
CRAX optimizes a density field $\hat{\rho}\in[0,1]$ on a 2D domain, and the density of each element sets its stiffness,
strength, and fracture toughness. The domain is either a finite structure with supports and a prescribed displacement, as
in standard fracture topology-optimization problems (\S\ref{sec:benchmarks}), or the periodic unit cell of an architected
material under a prescribed macroscopic strain (\S\ref{sec:validation}). Both share the pipeline of
Fig.~\ref{fig:framework}, shown there on a periodic cell, and its three components, summarized here and detailed in
Methods (\S\ref{sec:methods}).
\begin{figure}[p]
  \centering
  \includegraphics[width=\linewidth]{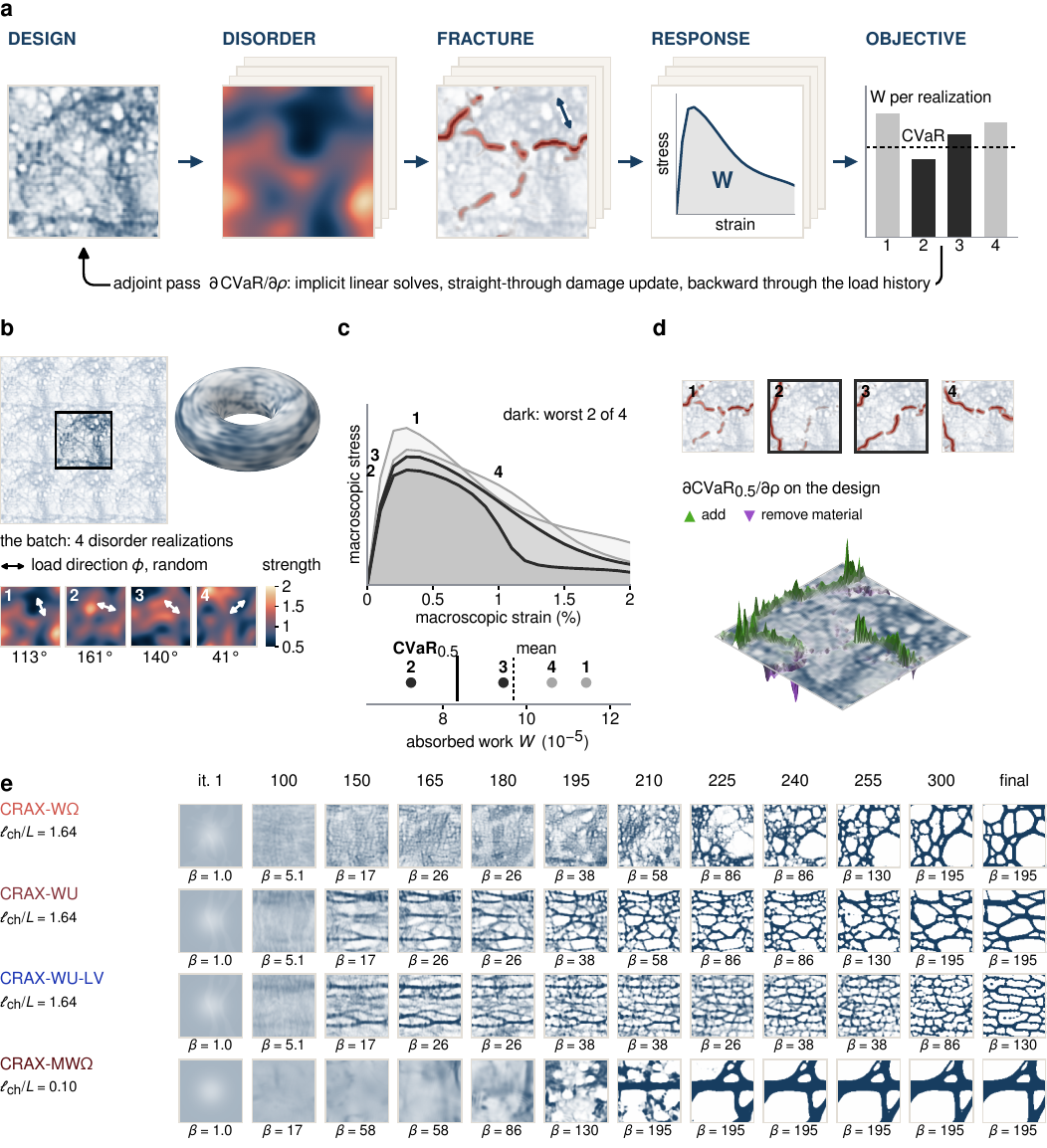}
  \caption{\textbf{Differentiable disorder-aware fracture topology optimization (CRAX).} Caption continues on the next page.}
  \label{fig:framework}
\end{figure}

\begin{figure}[p]
  \ContinuedFloat
  \begin{minipage}[t][0.75\textheight][t]{\linewidth}
  \caption*{\textbf{Figure~\ref{fig:framework} (continued). Differentiable disorder-aware fracture topology optimization (CRAX).}
  (a) Computational pipeline: a design density field, correlated strength and toughness fields, linear elastic finite elements coupled with phase-field cohesive damage, the
  absorbed work $W$ of each realization, and the \CVaR of the batch, with adjoint gradients flowing back through every load step.
  Tiles show one realization of a real optimization batch on an intermediate design termed "CRAX-W$\Omega$" (iteration 200); the
  arrow in the fracture tile is that realization's uniaxial load direction; dark bars, the worst two of the four realizations.
  (b) One periodic cell shown as a $3\times3$ tiling and on a torus; an evaluation batch's four material disorder field realizations, each loaded in uniaxial strain along its own random direction $\phi$ (white arrows); field statistics: Methods.
  (c) Stress--strain curves of the four realizations and each curve's corresponding work (shaded area under each curve; dark curves, the two lowest work values of the batch). $\CVaR_{0.5}$ is the mean of the worst two work values and is the main optimization objective used in this work.
  (d) The top four panels show the final damage (red) of the four realizations, overlaid atop physical density fields (dark frames denote the worst two, determining $\CVaR_{0.5}$). Bottom: gradient of $\CVaR_{0.5}$ with respect to the physical density of each element, from the exact adjoint (the mean of the gradients of realizations 2 and 3), drawn as an isometric relief over the physical density field (flat plane, white to dark blue, 0 to 1) where height is proportional to the scalar value of the local gradient: peaks (green) correspond to where adding material raises $\CVaR_{0.5}$ and pits (purple) correspond to where removing it does. Both additive and subtractive gradients localize near the cracks of the two worst realizations. Agreement of the exact adjoint and of the gradient used in optimization with central finite differences: SI, Fig.~S4, Section~S7.3, and Table~S10. (e) Physical density fields (white to dark blue, 0 to 1) of four CRAX designs during optimization given at several optimization iterations with the corresponding projection sharpness $\beta$. $\beta$ follows the gated continuation ($\times1.5$ per stage from 1; Methods). Optimization histories of $\beta$, intermediate-density fraction, and objective: SI, Fig.~S1.}
  \end{minipage}
\end{figure}

\emph{(i) A differentiable fracture solver.}
Material failure is modeled with the phase-field cohesive-zone model (\PFCZM)~\cite{wu_2017_Unified, wu_2018_Length}, which regularizes a cohesive crack of strength $f_t$ and toughness $G_c$ over a length $\ell$ and, unlike earlier brittle phase-field variants, has a failure strength and traction--separation law that does not depend on $\ell$ (\S\ref{sec:pfczm}). The load is raised
step by step; damage accumulates, degrades the local stiffness, and redistributes stress, so the response follows crack
initiation, propagation, and the loss of load-bearing capacity. The pipeline is differentiable end to end (Fig.~\ref{fig:framework}a). Rather than unrolling solver iterations, which is memory-intensive and can produce incorrect gradients when discrete constraint conditions are recalculated (\S\ref{sec:diff}), we differentiate each solve implicitly at its converged state.

For the bound-constrained damage problem we implement a
semi-smooth adjoint on the converged active set (\S\ref{sec:diff}; SI, Section~S1.9). The designs reported in this work were optimized with straight-through estimators in
place of this exact adjoint; their gradient pointed uphill at four of the five design states tested (\S\ref{sec:diff}; SI, Section~S7.4). The exact adjoint, which matches central finite differences within 2\,\% (SI, Fig.~S4), served for verification and for a re-optimization of one design, which changed the cell about as much as independent runs differ but not its certified performance (SI, Section~S13); every periodic cell is evaluated separately from its optimization (\S\ref{sec:validation}).

\emph{(ii) Correlated material disorder.}
Empirically, brittle materials fail where correlated weak regions meet stress concentrations, not at their mean strengths. Therefore every design is endowed with multiplicative lognormal strength and toughness fields with Gaussian autocorrelation (correlation length 0.15 times the width of the domain, cell, or structure; coefficient of variation $0.25$; cross-correlation $0.75$), sampled spectrally and, on the periodic cell, seamless across the cell boundary (\S\ref{sec:fields}). These material property fields are stochastically varied for each forward-simulation used to compute gradients during TO. We denote one realization of these fields by $\xi$. For the cells designed for unknown load directions (CRAX-W$\Omega$ and CRAX-MW$\Omega$), each realization also carries its own uniformly random load direction $\phi$, measured from the cell $x$ axis (Fig.~\ref{fig:framework}b).

\emph{(iii) A tail-risk objective.}
We maximize the conditional value-at-risk of the absorbed work $W$, the mean
of the worse half of the realizations in each stochastic batch (the worst two of four for the periodic cells; the worse of two for the standard problems),
\begin{equation}\label{eq:cvar_main}
  \max_{\rho}\; \widehat{\CVaR}_{0.5}\!\left[W(\rho,\xi_1),\dots,W(\rho,\xi_k)\right]
  \;=\; \max_{\rho}\; \frac{2}{k}\sum_{s\,\in\,\mathcal{K}(\rho)} W(\rho,\xi_s),
\end{equation}
where the hat marks the estimate from one batch of $k$ realizations $\xi_1,\dots,\xi_k$ and $\mathcal{K}(\rho)$ is the
set of the $k/2$ with the lowest work, so that each design is graded on its unlucky tail rather than its average
(\S\ref{sec:objective}). The gradient of this objective, shown from the exact adjoint in Fig.~\ref{fig:framework}d, concentrates near the cracks of the two worst realizations of the batch (Fig.~\ref{fig:framework}c).

Four CRAX cells are followed below (Fig.~\ref{fig:framework}e), each emerging from a nearly uniform density as the projection sharpens. Their names give the recipe: W, optimized for absorbed work; U, uniaxial strain along $x$; $\Omega$, random load directions; LV, a local-volume constraint that excludes large voids. Brittleness is set by the ratio of the material's Irwin characteristic length, $\ell_{ch} = E G_c/f_t^2$ ($E$, Young's modulus), to the cell size $L$ (smaller is more brittle, equivalent to a larger cell of the same material): three cells were optimized at $\ell_{ch}/L = 1.64$, and CRAX-MW$\Omega$ (M, macro: a larger cell of the same material) is the CRAX-W$\Omega$ recipe at 0.10, which gave a
much coarser cell.

\subsection{Standard fracture problems}\label{sec:benchmarks}

Before designing materials, we tested the solver and the objective on finite structure s that the fracture topology-optimization literature already knows: eight well-studied fracture problems with standard supports and a prescribed
displacement (Fig.~\ref{fig:certification}a,b; \S\ref{sec:finite}). Run in plane stress, as Wu and Nguyen~\cite{wu_2018_Length} assume, the solver reproduces their peak loads within about 2\,\% for single-edge-notched tension (SENT), SENT shear, and notched three-point bending (SI, Section~S7.1, Table~S8, and Fig.~S3a). We then optimized each problem for the worse of two disorder realizations per iteration and compared the result with its stiffness-optimal twin at equal volume (35\,\%), on a deterministic field and on four disordered fields (Fig.~\ref{fig:certification}a,b). The fracture-optimized twin absorbed more work on every field of seven of the eight problems, 1.31 to 4.10 times the work of its stiffness twin (geometric means over the four fields; Fig.~\ref{fig:certification}b; SI, Table~S9; twin pairs failing: Supplementary Movies S1--S10 and SI, Section~S16). Mixed-mode SENT is a tie: the gain held on only two of the four fields. In V-notch bending, the fracture-optimized twin carries 3.8 times the peak load and reaches its peak at 2.6 times the displacement of its stiffness twin in the deterministic field (SI, Fig.~S3b): it delays failure rather than merely stiffening the load path.

\subsection{Architected materials under uniaxial strain applied in one direction}\label{sec:validation}\label{sec:canonical}
We then turned to the intended use, the periodic cell of an architected material, judged against stiffness-optimal and canonical cells. The material is modeled as one periodic unit cell loaded by a prescribed macroscopic strain $\epsbar$ through
the split $u(X) = \epsbar\,X + \tilde{u}(X)$, where the periodic fluctuation $\tilde{u}$ carries zero average strain; the
macroscopic stress follows from the damaged stiffness in a Hill--Mandel consistent manner~\cite{hill_1963_Elastic} (\S\ref{sec:homog}). The cell is assembled on a torus, so periodicity needs no constraint
equations and cracks cross the cell boundary freely (Fig.~\ref{fig:framework}b). Designs are \emph{discovered} under an
economical discretization whose biases we characterize in the SI (Section~S5 and Fig.~S2) and \emph{evaluated} under a
converged, pre-specified protocol on held-out disorder fields shared by competing designs.

Our references (Fig.~\ref{fig:certification}c) are stiffness-optimized cells (SIMP-KU, SIMP-KU-LV, and SIMP-K$\Omega$) and
canonical/bioinspired cells (running bond, squashed honeycomb, regular hexagon, and a Voronoi foam), closely volume-matched in physical density (SI, Table~S3). We first asked whether the two uniaxially work-optimized cells, CRAX-WU and CRAX-WU-LV, outperform them along their optimization direction. We compare work per unit solid volume (specific work; \S\ref{sec:canonical_methods}).

CRAX-WU absorbed more work than SIMP-KU-LV and SIMP-K$\Omega$, by 26.1\,\% and 44.6\,\%, and, in the lower tail, more than
SIMP-KU, the stiffness-optimal cell for the same loading (dots with bars in Fig.~\ref{fig:certification}d; SI, Table~S6). Against SIMP-KU, the 10.8\,\% gain in mean work was not resolved, but $\CVaR_{0.5}$, the mean work of the worse half of the realizations, was 23.5\,\% higher. SIMP-KU reached about 1.5 times the peak stress of
CRAX-WU on average and a higher peak across all 48 realizations (Fig.~\ref{fig:certification}d,e), which suggests that the fracture objective traded peak stress for the absorbed work it rewards.

Among the canonical cells,
CRAX-WU exceeded the regular hexagon and the Voronoi foam in work and tied running bond and the squashed honeycomb, with lower tails within 5\,\% of both. It exceeded every canonical cell in peak stress at $\phi = 0$, by 28\,\% (squashed honeycomb) to 109\,\% (running bond
loaded \textit{across} its bricks; Fig.~\ref{fig:certification}d; SI, Table~S6). Loaded instead \textit{along} its bricks ($\phi = 90^\circ$), running bond reached 1.10 times the peak stress of CRAX-WU on the same
48 realizations, with a difference in work that was not statistically resolved (0.94 times; Fig.~\ref{fig:certification}d,e; SI,
Table~S6). CRAX-WU thus lies on the work--strength Pareto front of the tested cells, with SIMP-KU, the squashed honeycomb, and running bond loaded along its bricks (Fig.~\ref{fig:certification}e); from $2\times2$ supercells upwards, however, running bond absorbed more work per cell than CRAX-WU and CRAX-WU-LV (SI, Section~S8).

CRAX-WU-LV, optimized with a local-volume constraint to avoid large voids, tied CRAX-WU in work and gave the same mean-work comparisons within two
percentage points (open diamonds in Fig.~\ref{fig:certification}d; SI, Table~S5), absorbing 25\,\% more than SIMP-KU-LV, the stiffness-optimal cell under the same constraint. The constraint also left CRAX-WU-LV without fragile sites under
single-member loss (\S\ref{sec:mechanisms}), at no cost in
work along the optimization direction and 25\,\% in peak stress (SI, Section~S6.1 and Tables~S5 and S6). Work ratings of single cells change with domain size, whereas strength ratings barely do (SI, Section~S8 and Fig.~S6).

\begin{figure}[p]
  \centering
  \includegraphics[width=\linewidth]{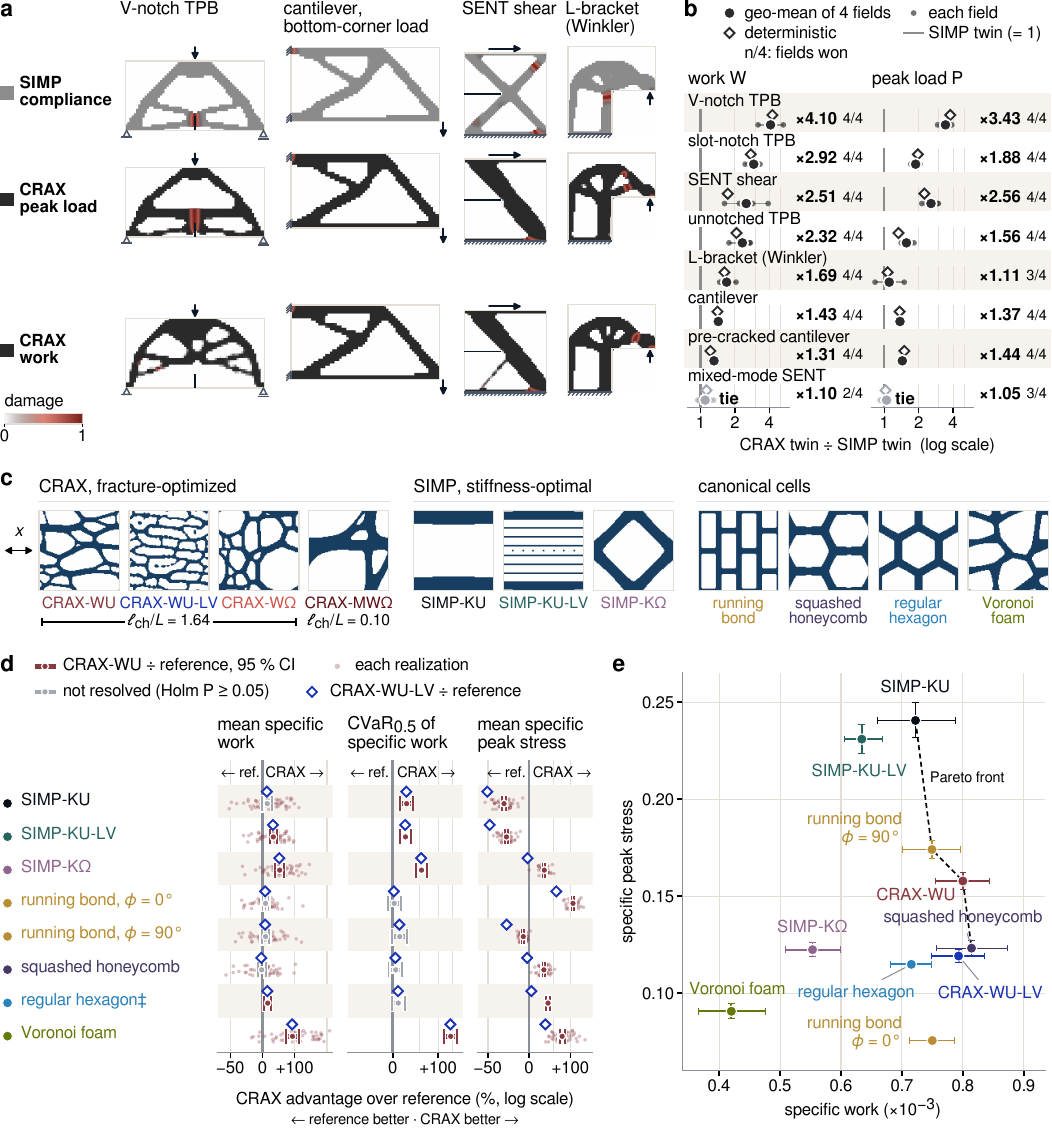}
  \caption{\textbf{Fracture-optimized designs absorb more work than their stiffness-optimal counterparts, on standard
  fracture problems and, at least in the lower tail, as periodic cells, and match the best canonical cells in work, with CRAX-WU on the work--strength Pareto front.} Caption continues on the next page.}
  \label{fig:certification}\label{fig:paired}\label{fig:canonical}
\end{figure}

\begin{figure}[p]
  \ContinuedFloat
  \begin{minipage}[t][0.75\textheight][t]{\linewidth}
  \caption*{\textbf{Figure~\ref{fig:certification} (continued). Fracture-optimized designs absorb more work than their stiffness-optimal counterparts, on standard
  fracture problems and, at least in the lower tail, as periodic cells, and match the best canonical cells in work, with CRAX-WU on the work--strength Pareto front.}
  (a) Stiffness-optimal designs from SIMP compliance minimization (top row) and CRAX twins optimized for peak load
  (middle) and for absorbed work (bottom) on four of the eight standard problems at 35\,\% volume, with the final damage
  in the deterministic field; for each objective the start with the higher work ratio is shown, as in (b) (starts and ratios: SI, Section~S7.2).
  (b) CRAX twin divided by stiffness twin for work and peak load on every standard problem (same problem, volume, and
  fields; log scale; best of four variants per problem, all in SI, Section~S7.2); symbols as in the key. Solver verification against Wu and Nguyen and the V-notch load paths: SI, Fig.~S3.
  (c) One periodic cell of each design, binarized as evaluated (solid fractions: SI, Table~S3); $\ell_{ch}/L$, characteristic length used
  in optimization over the cell size (CRAX-W$\Omega$ is compared only over load directions; CRAX-MW$\Omega$ is not compared in d and e). The regular hexagon is shown as a square window of its rectangular
  ($1\times\sqrt3$) cell; $x$, the strain direction in (d) and (e). Each design keeps its color in every figure.
  (d) Advantage of CRAX-WU over each reference under uniaxial strain along $x$, on the same 48 held-out disorder
  realizations, in percent on log scales: mean specific work, $\CVaR_{0.5}$ of specific work
  (mean of the worse 24 realizations) and mean specific peak stress. Dots with bars, ratio of means and 95\,\%
  paired-bootstrap CI (not adjusted for multiplicity; the peak-stress intervals, $\pm$2--4\,\%, are about as wide as the
  dot); other symbols as in the key. Running bond appears loaded across its bricks ($\phi = 0$) and along them ($\phi = 90^\circ$). Grey markers indicate Holm-adjusted two-sided sign-flip $P \geq 0.05$ across the eight references. $^\ddagger$Evaluated on its own 48
  realizations (unpaired bootstrap and permutation test). Values and $P$: SI, Tables~S5 (CRAX-WU-LV) and S6 (CRAX-WU).
  (e) Specific peak stress against specific work under uniaxial strain along $x$: mean of 48 realizations with 95\,\%
  bootstrap CIs; running bond in both directions, as in (d); dashed line, Pareto front of the plotted designs (SIMP-KU, running bond at $\phi = 90^\circ$, CRAX-WU, squashed honeycomb). Specific
  quantities are per unit solid volume, in code units.}
  \end{minipage}
\end{figure}

\subsection{Architected materials under uniaxial strain applied in unknown directions}\label{sec:direction}
Thus far every cell we reported was judged along one direction. We next asked how the cells perform when the load direction is not known a priori. Fig.~\ref{fig:direction}a plots each property against load direction over the rose (the 12 load directions, $0^\circ$ to $165^\circ$ in $15^\circ$ steps, named after the polar rose diagrams in which such data are drawn). Direction dependence is measured by $D$, the noise-corrected standard deviation of the log direction means over the rose, which approximates the fractional variation between directions and is 0 when properties among different directions vary no
more than properties among realizations (replicates) do (\S\ref{sec:exp_stats}).

In Fig.~\ref{fig:direction}b each cell is one point, placed by its $D$ in stiffness, strength, and work. Taking the three properties together, CRAX-W$\Omega$, the design optimized over random load
directions, had the lowest maximum direction dependence of any cell: $D = 0.075$ in Young's modulus, 0.02 in peak stress, and 0.10 in work, simulated at $\ell_{ch}/L = 1.64$. Every other cell reached
$D \geq 0.19$ in at least one of the three: the hexagon and the honeycomb in work (0.19 and 0.20), and CRAX-WU-LV, running
bond, SIMP-K$\Omega$, CRAX-WU, CRAX-MW$\Omega$, and the Voronoi foam in stiffness (SI, Table~S38, which also lists $D$
at a more brittle setting). Much of the simulated work anisotropy of the hexagon is a discretization effect, however, as its $\pm30^\circ$ walls are staircased on the square mesh, and its symmetry-equivalent directions differ by up to a factor of 1.76 in work (SI, Section~S9; its prints: \S\ref{sec:printed}). In stiffness alone, only the hexagon, isotropic by its six-fold symmetry, was more uniform
than CRAX-W$\Omega$ ($D = 0.004$), and three further runs of the CRAX-W$\Omega$ recipe reached $D$ of 0.13--0.15 in work
and at most 0.04 in peak stress (SI, Section~S11.1). The walls of CRAX-W$\Omega$ face every direction almost equally (Fig.~\ref{fig:novelty}f; \S\ref{sec:mechanisms}).

Property by property, the failure of CRAX-W$\Omega$ depended little on direction (Fig.~\ref{fig:direction}a). Its peak
stress hardly varied, and in specific work its best and worst direction means differed by a factor of 1.40, with about
half the direction dependence of the squashed honeycomb and the regular hexagon, whose peak stresses, like that of CRAX-W$\Omega$, hardly varied with direction. Running bond varied little in work ($D = 0.05$) but 3.3-fold in peak stress ($D = 0.43$) and strongly in stiffness ($D = 0.79$). The random Voronoi foam depended more on direction than CRAX-W$\Omega$ in both stiffness and work ($D = 0.30$
against 0.075, and 0.22 against 0.10; SI, Table~S38).

\begin{figure}[p]
  \centering
  \includegraphics[width=\linewidth]{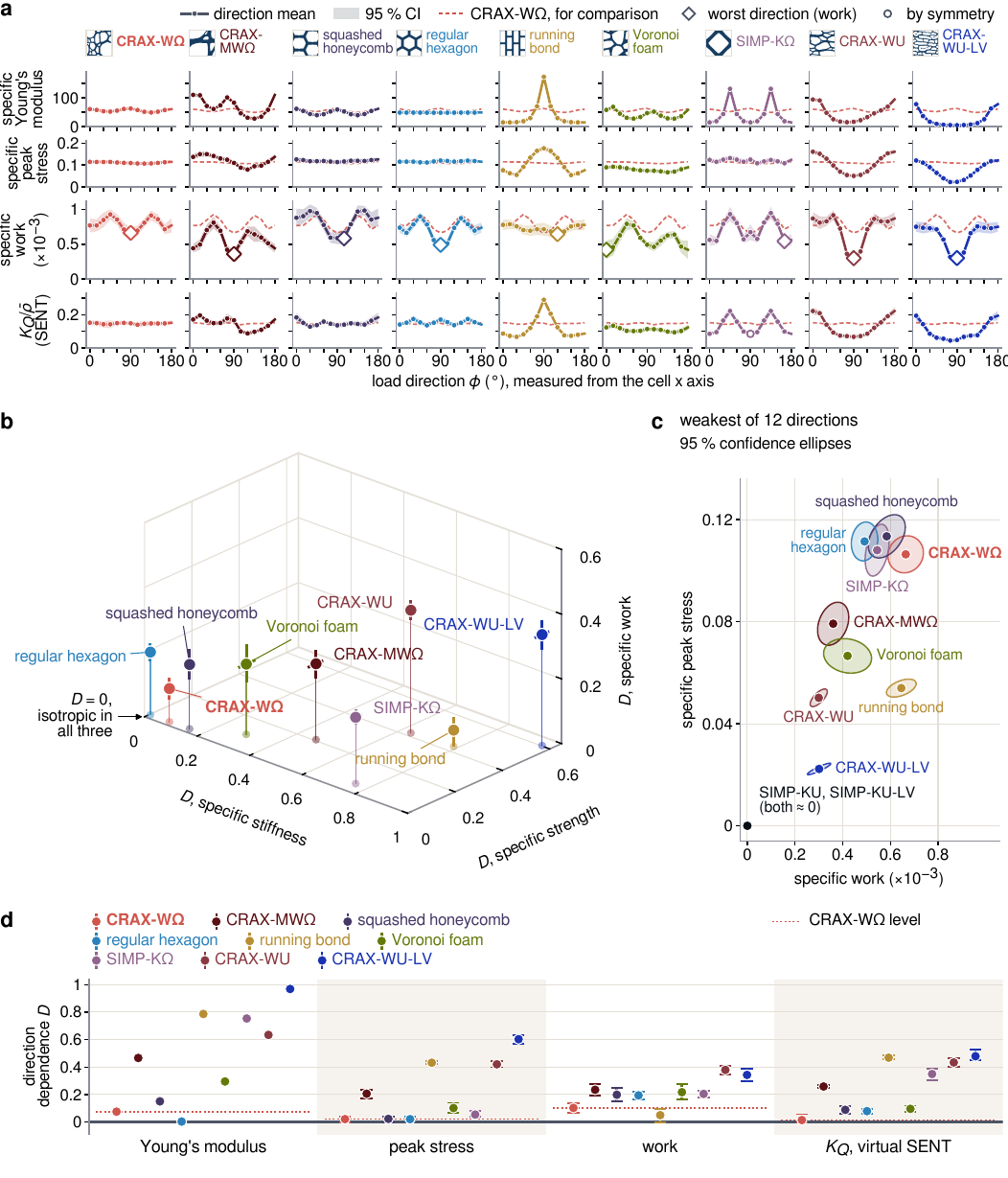}
  \caption{\textbf{A cell optimized over random load directions has the lowest maximum direction dependence of simulated stiffness, strength, and work among the tested cells.} Caption continues on the next page.}
  \label{fig:direction}
\end{figure}

\begin{figure}[p]
  \ContinuedFloat
  \begin{minipage}[t][0.75\textheight][t]{\linewidth}
  \caption*{\textbf{Figure~\ref{fig:direction} (continued). A cell optimized over random load directions has the lowest maximum direction dependence of simulated stiffness, strength, and work among the tested cells.}
  (a) Specific Young's modulus, specific peak stress, and specific work of the periodic cell, and specific apparent
  toughness $K_Q/\bar\rho$ ($\bar\rho$, solid fraction of the cell) of virtual SENT specimens ($8\times9$ cells, crack 0.45 of the width; SI, Section~S9) against the
  direction $\phi$ of uniaxial strain from the cell $x$ axis, for the nine cells drawn above the columns (hexagon as in Fig.~\ref{fig:certification}c); CRAX-MW$\Omega$, optimized at $\ell_{ch}/L = 0.10$, is evaluated here at 1.64. Lines, direction means ($n = 16$ held-out
  realizations for peak stress and work, $n = 4$ for $K_Q$, paired across the square-cell designs; Young's modulus is deterministic);
  bands, 95\,\% $t$ intervals; dashed lines, CRAX-W$\Omega$; open diamonds, worst direction of work; open circle,
  $K_Q/\bar\rho$ of SIMP-K$\Omega$ at $90^\circ$, equal to $0^\circ$ by symmetry; the point at $180^\circ$ repeats $0^\circ$. Ratios of the
  largest to the smallest direction mean: SI, Table~S14 (work); direction dependence $D$ (Methods; 0 when directions differ no
  more than realizations): SI, Table~S38 and (d).
  (b) Direction dependence $D$ of specific stiffness, specific strength (peak stress) and specific work of each cell,
  simulated at $\ell_{ch}/L = 1.64$ (orthographic projection, equal scale per unit $D$): 95\,\%
  bias-centred bootstrap intervals for strength and work (stiffness is deterministic); drop lines to the stiffness--strength plane. The printed lattices:
  Fig.~\ref{fig:experiments}g. SIMP-KU and SIMP-KU-LV carry essentially no load transverse to their struts (direction
  mean $\approx 0$ at $90^\circ$), so their $D$ is undefined; they appear only in (c), at the origin.
  (c) Weakest of the 12 load directions ($0^\circ$ to $165^\circ$ in $15^\circ$ steps, 16 realizations per direction): minimum
  of the direction means, taken separately for work and peak stress. Ellipses, 95\,\% confidence regions from 10{,}000
  bootstrap resamples of the 16 realizations (jointly across directions) with the weakest direction re-selected in each
  resample ($\chi^2_2$ region of the bootstrap covariance, centred on the estimate).
  (d) $D$ of each property for every cell with 12-direction roses, with 95\,\% bias-centred bootstrap intervals (grey bar
  for Young's modulus: range over three mesh refinements; 0, isotropic); dotted lines, the CRAX-W$\Omega$ level.
  Omnidirectional statistics of CRAX-W$\Omega$ against every design: SI, Fig.~S7. Specific quantities are per unit solid
  volume.}
  \end{minipage}
\end{figure}

Virtual SENT specimens of $8\times9$ cells, cracked to 0.45 of their width and cut at the 12 lattice angles (four
realizations each; geometry and loading in SI, Fig.~S9a), extended the comparison to cracks (Fig.~\ref{fig:direction}a, bottom row, and d). Their \textit{apparent toughness} $K_Q$, the
peak load scaled by an isotropic geometry factor, belongs to a specimen with a 4.4-cell ligament; it technically is not a material
fracture toughness $K_{\mathrm{IC}}$. CRAX-W$\Omega$ had the flattest $K_Q$ rose of the nine designs with a transverse load path and a lower direction dependence than each of the five references ($D = 0.01$ against 0.08--0.47; Holm-adjusted one-sided $p \leq 0.03$; SI, Section~S9 and Table~S17). In specific apparent toughness $K_Q/\bar\rho$, its worst direction exceeded those of the Voronoi foam, SIMP-K$\Omega$, and running bond by 48--109\,\% and tied those of the honeycomb and the hexagon. With $K_Q$ as a fourth property, the largest $D$ of CRAX-W$\Omega$ stayed at 0.10 (Fig.~\ref{fig:direction}d). Wider virtual compact-tension specimens are reported in SI, Section~S9 and Fig.~S9b--d.

We then asked whether this uniformity costs omnidirectional performance, judging CRAX-W$\Omega$ on three statistics: the
mean over all directions and realizations, the mean in the worst direction, and $\CVaR_{0.5}$ over all 192 simulations of
the periodic cell (12 directions $\times$ 16 realizations; Fig.~\ref{fig:direction}c; SI, Section~S9, Fig.~S8, and
Tables~S14 and S16). CRAX-W$\Omega$ exceeded SIMP-K$\Omega$ on all three statistics, by 11.5--22\,\%, and more than doubled the
worst-direction work of CRAX-WU-LV and CRAX-WU (SI, Table~S15 and Fig.~S7). Running bond and the squashed honeycomb
remained competitive: CRAX-W$\Omega$ had 10\,\% more mean work than running bond but tied it in the worst direction and in
the tail, and it tied the honeycomb on all three statistics. CRAX-W$\Omega$ exceeded the regular hexagon on all three statistics,
by 12--35\,\%. CRAX-W$\Omega$ ties for the best of four runs of its recipe: two further seeds absorbed 8--10\,\% less mean work and only tied the regular hexagon, running bond, and
SIMP-K$\Omega$ (SI, Section~S11.1). For this recipe, the run reported here thus combined the lowest maximum direction dependence of the tested cells with the
omnidirectional performance of the best canonical lattices.

\subsection{3D-printed architected dogbones in tension}\label{sec:printed}
We printed CRAX-WU, CRAX-W$\Omega$, CRAX-MW$\Omega$, the squashed honeycomb, the regular hexagon, running bond, and a solid reference as
ASTM D638-14 Type~I dogbones by masked stereolithography, and loaded them in tension (Fig.~\ref{fig:experiments}a; Methods). The resin's measured modulus and strength, with a literature toughness of comparable photopolymers, give a characteristic
length about 40 times shorter relative to the cell than the one used in optimization ($\ell_{ch}/L \approx 0.04$; Methods),
so the prints were far more brittle than their simulated counterparts. The tests therefore ask whether the direction dependence and the overall ordering of the designs carry over to a far more brittle setting, not whether the simulated magnitudes do. We tested 77 specimens, two or three per design and
direction, and report values per unit designed solid mass.

Loaded along its optimization direction ($\phi = 0$, $n = 3$; work, $n = 2$), CRAX-WU reached a specific peak stress of 22.4~kN\,m\,kg$^{-1}$ and a specific work of 450~J\,kg$^{-1}$ (Fig.~\ref{fig:experiments}b,c; SI, Table~S34). Running bond, printed only with its bricks along the load ($\phi = 90^\circ$, $n = 2$),
reached 1.10 times the specific peak stress of CRAX-WU and 0.85 times its specific work, and the two could not be statistically distinguished (SI, Table~S35, Section~S18.3). In the plane of specific work and specific peak stress, running bond along its bricks and CRAX-WU formed the Pareto front of the printed lattices (Fig.~\ref{fig:experiments}c), as they do among the simulated cells (Fig.~\ref{fig:certification}e). The
squashed honeycomb at $\phi = 0$, its strongest printed direction, fell below CRAX-WU in specific peak stress (0.78 times, resolved) and in specific work (0.82 times, not resolved; SI, Section~S18.5).
The dense solid exceeded every lattice of Fig.~\ref{fig:experiments}c per unit mass, in specific peak stress as the scaling of cellular solids predicts~\cite{gibson_1997_Cellular} and in specific work as well (SI, Section~S18.5 and Table~S34); it serves only as a reference.

CRAX-W$\Omega$, optimized over random load directions, reached about half the specific peak stress and work of CRAX-WU along $\phi = 0$ (11.9~kN\,m\,kg$^{-1}$ and 204~J\,kg$^{-1}$; Fig.~\ref{fig:experiments}b,c), but its direction means over
all 12 printed directions stayed between 11.9 and 15.1~kN\,m\,kg$^{-1}$, meaning that CRAX-W$\Omega$ trades peak performance along any one direction for a strength and work that hardly vary with direction. It was printed in 12 directions (with replicates) and compared with the two lattices whose printed directions cover
the full rose by symmetry, the squashed honeycomb (four directions) and the regular hexagon (three directions;
Fig.~\ref{fig:experiments}e,f). Averaged over all directions, the three designs could not be distinguished in specific peak stress or work, but they differed in directional variance. CRAX-W$\Omega$ reached 13.4~kN\,m\,kg$^{-1}$ and 233~J\,kg$^{-1}$, 0.90 to 1.02 times the specific peak stress and work of the two lattices (SI, Tables~S34 and S35). In specific peak stress, the direction dependence $D$ (\S\ref{sec:direction}) was 0.03 for CRAX-W$\Omega$, 0.20 for the honeycomb, and 0.13 for the hexagon (SI, Table~S38). In specific work, the direction means of CRAX-W$\Omega$ varied no more than its replicates ($D$ estimated at 0, 95\,\% interval 0--0.10), against 0.40 for the honeycomb and 0.30 for the hexagon (Holm-adjusted $p = 0.024$ and 0.046). With two or three specimens per direction, the test would detect only $D \geq 0.18$ (80\,\% power), so these contrasts indicate less direction dependence than the two lattices, not perfect isotropy (SI, Section~S18.5 and Fig.~S22).
Moreover, stiffness isotropy did not imply failure isotropy (Fig.~\ref{fig:experiments}g). The printed hexagon had a flat stiffness rose (specific elastic modulus, max/min 1.05; SI, Table~S37), yet its specific work varied 2.3-fold with direction (its print mask keeps the staircased raster of the $\pm30^\circ$ walls, and printing adds direction-dependent effects of its own; SI, Sections~S9 and S18.1), and it matched the anisotropic honeycomb in rose-mean specific
peak stress and work (Fig.~\ref{fig:experiments}f; SI, Table~S35).

In its weakest printed direction ($\phi = 15^\circ$, $n = 2$), CRAX-W$\Omega$ absorbed 1.46 and 1.32 times the specific
work of the honeycomb and the hexagon in theirs, and specific peak stress ranked the designs alike, but neither advantage
was resolved, and taking the minimum over 12 directions rather than four or three biases these ratios against
CRAX-W$\Omega$ (SI, Table~S35). A test-day adjustment changed the lattice ratios above by at most 0.06 (Methods; SI,
Section~S18.5).

Printed CRAX-W$\Omega$ broke on average perpendicular to
the load, with no dependence on direction (mean angle $-2^\circ$, s.d.\ $18^\circ$, 22 specimens;
$p = 0.13$; Fig.~\ref{fig:experiments}a,d), and of the printed specimens with video, four of ten
honeycombs fragmented and none of 24 CRAX-W$\Omega$ (SI, Section~S10.3).

CRAX-MW$\Omega$ (the CRAX-W$\Omega$ recipe at $\ell_{ch}/L = 0.10$, printed in four directions) varied little across strain directions in specific work ($D = 0$) but strongly in specific elastic modulus ($D = 0.53$) and peak stress ($D = 0.26$), properties its recipe did not target (SI, Section~S18.5 and Table~S39).

In stiffness and in work, the printed $D$ values fell in the simulated order of the three designs, and in peak stress,
where the simulated cells were tied ($D \approx 0.02$ each), the prints separated them. However, CRAX-W$\Omega$
was not resolved from either lattice in stiffness (SI, Section~S18.5 and Tables~S36--S38). Finite-dogbone simulations at $\ell_{ch}/L = 0.10$, used only as rankings, ordered the 14 simulated design--directions
largely as measured (Spearman $\rho = 0.85$ for specific peak stress and 0.78 for specific work; SI, Section~S18.5).

\begin{figure}[p]
  \centering
  \includegraphics[width=\linewidth]{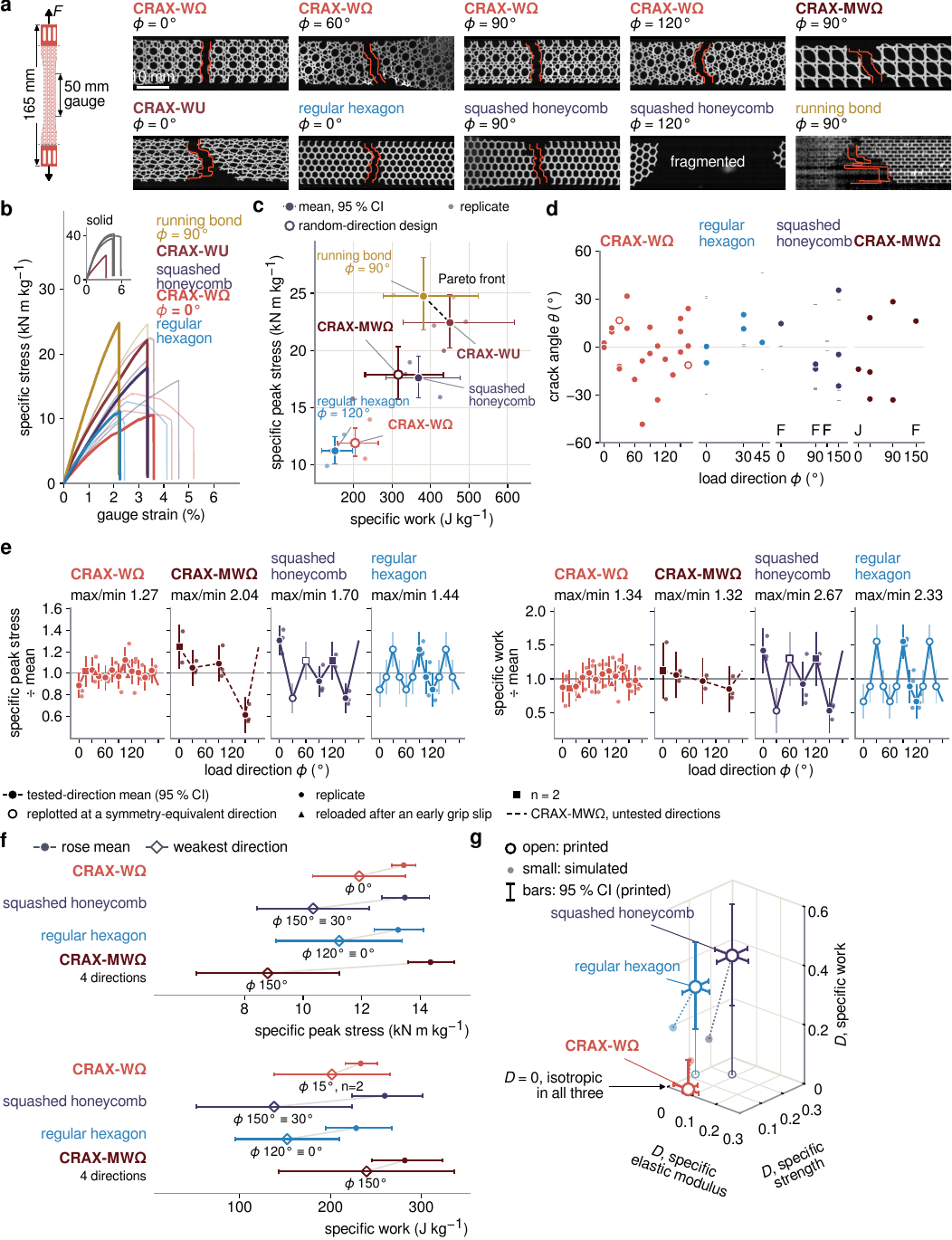}
  \caption{\textbf{Printed in a far more brittle regime than used in optimization, CRAX-W$\Omega$ shows no detectable direction dependence of strength or work over 12 load directions.} Caption continues on the next page.}
  \label{fig:experiments}
\end{figure}

\begin{figure}[p]
  \ContinuedFloat
  \begin{minipage}[t][0.75\textheight][t]{\linewidth}
  \captionsetup{singlelinecheck=false}
  \caption*{\textbf{Figure~\ref{fig:experiments} (continued). Printed in a far more brittle regime than used in optimization, CRAX-W$\Omega$ shows no detectable direction dependence of strength or work over 12 load directions.}
  (a) The CRAX-W$\Omega$ dogbone and the last video frame of ten printed dogbones at fracture (load axis horizontal;
  $\pm24$~mm about the crack) with the faces of the two fragments (red); $\phi$, angle between the load and the cell $x$
  axis; $\theta$, crack angle (0, perpendicular to the load; plotted in d); the hexagon frame (printed at $\phi = 120^\circ$) is labeled with $\phi$ modulo $60^\circ$. Frames use a fixed
  grey-level map per recording (Data availability); the underexposed running-bond crack ran in the fillet next to the
  grip and stepped along the brick joints; the fragmented honeycomb lost the material between its grips.
  (b) Specific stress against gauge strain (video; Methods; the CRAX-WU test with peak force only is not drawn) for
  CRAX-WU, CRAX-W$\Omega$, and the squashed honeycomb at $\phi = 0$, the regular hexagon at $\phi = 120^\circ \equiv 0^\circ$ and
  running bond at $\phi = 90^\circ$, its only printed direction; bold,
  the replicate closest to the group median; inset, the vertically printed solid (with a CRAX-WU curve for scale).
  (c) Specific peak stress against specific work at $\phi = 0$ (hexagon, $120^\circ$; running bond, $90^\circ$): raw means
  (not day-adjusted) with 95\,\% confidence intervals in both quantities (large dots) and replicates with both values (small dots); $n = 3$ (peak) and 2 (work) for
  CRAX-WU, 2 and 2 for CRAX-MW$\Omega$ and running bond, 3 and 3 for CRAX-W$\Omega$, the honeycomb, and the hexagon; dashed line, Pareto front of the lattice means (running bond at $\phi = 90^\circ$ and CRAX-WU, whose
  difference is not resolved). The vertically printed solid is not shown (SI, Table~S34). Ratios to CRAX-WU: text and SI, Section~S18.5.
  (d) Printed crack angle against load direction $\phi$: filled dots, single crack; open dots, recording ends before separation;
  F, fragmented; J, break inside a grip; short grey lines, cell-row directions measured in the frame; the hexagon is plotted
  at $\phi$ modulo $60^\circ$. For CRAX-W$\Omega$ (22 specimens) $\theta$ does not depend on $\phi$ (harmonic regression,
  $p = 0.13$).
  (e) Direction means of specific peak stress and specific work over the design's mean: tested directions
  (filled, solid lines) with 95\,\% one-way pooled-variance $t$ intervals; other symbols as in the key; CRAX-MW$\Omega$ over the mean of its four printed directions; $n = 3$ per direction, squares $n = 2$; max/min, largest over smallest direction mean.
  (f) Rose mean (filled; interval as in c) and weakest printed direction (open; interval as in e) of each design; ratios
  of CRAX-W$\Omega$ to each design and of the hexagon to the honeycomb, with the ratios expected for equal, isotropic designs:
  SI, Table~S35. CRAX-MW$\Omega$: over its four
  printed directions.
  (g) Direction dependence $D$ of the printed lattices (day-adjusted primary analyses; Methods) in specific elastic modulus,
  specific peak stress, and specific work, with 95\,\% intervals (open markers; equal scale per unit $D$); small translucent markers, the same designs simulated at the brittleness used in optimization, joined by
  dotted lines; thin vertical lines drop to the elastic-modulus--strength plane. CRAX-MW$\Omega$ (four directions): SI, Table~S39.
  Intervals in (c) and rose-mean intervals in (f): pooled-variance $t$ intervals around the raw means (Methods).}
  \end{minipage}
\end{figure}

\subsection{Motif classes without a template}\label{sec:novelty}

No CRAX structure was generated based on a biological or canonical geometry: every design emerged from a nearly uniform density (\S\ref{sec:param}), except for three runs warm-started from CRAX-W$\Omega$ (SI, Section~S17.1). To ask how
the resulting cells relate to biological materials~\cite{wegst_2015_Bioinspired}, we compared all twelve CRAX designs (the three main-text designs optimized at $\ell_{ch}/L = 1.64$, CRAX-MW$\Omega$, and eight further SI runs) with seven canonical motifs, six of them bioinspired and one triangulated engineering reference (SI, Section~S17.1 and Fig.~S18), and with 240 random structures at solid fraction 0.350 from six families of foams, labyrinths,
jittered running bonds, and strut networks (Fig.~\ref{fig:novelty}a; SI, Section~S17.1). In brief, the designs fall into motif classes that biological and
canonical cellular solids also use: the closed-cell, trivalent, polydisperse foam (CRAX-W$\Omega$, CRAX-WU) and, for
CRAX-WU-LV, a lamellar and anisotropic variant (SI, Table~S28) that raters read as staggered, nacre-like brickwork; no instrument shows that a design copies a particular biological structure.

Twenty-one size-free descriptors from cellular-solid and bone morphometry (Fig.~\ref{fig:novelty}c; SI, Section~S17.2 and Table~S27) show that the
three main-text designs are closed-cell, predominantly trivalent, and polydisperse, with larger pores having more neighbors, consistent with Lewis's law~\cite{lewis_1928_Correlation} (SI, Table~S28): the topology of the closed-cell
foams that natural and engineered cellular solids share~\cite{gibson_1997_Cellular}. CRAX-W$\Omega$ is strongly polydisperse even among random bimodal foams (pore-diameter CV 0.75 and
topological disorder $\mu_2 = 6.8$~\cite{weaire_1984_Soap}, at their 82.5th and 95th percentiles); of the canonical motifs, the glass-sponge lattice matches it in pore-size polydispersity and topological disorder, though only in these two of the 21 descriptors.
What most consistently separates the designs from random foams is strut thickness: its between-strut coefficient of variation lies at the 97.5th percentile or above in the nearest random family of each main-text design (SI, Fig.~S17b and Table~S28), and nine of the twelve CRAX designs lie above the 95th
percentile. Since the random foams were built with walls of nearly uniform thickness, this result shows that the optimizer
places material unevenly.

Three blind raters, each a large-language-model instance without study context (Claude Fable 5.1), labeled 36 anonymized renders in independent random orders:
seven CRAX designs, two SIMP baselines, seven motifs, and 20 random structures (SI, Fig.~S17c and Section~S17.4; Fleiss
$\kappa = 0.70$ for the biological match, 0.61 for the origin). The labels follow the
morphological class: CRAX-W$\Omega$ and its version re-optimized at the certification discretization (SI,
Section~S11.2) are ``closed-cell foam (cork)'' in 6 of 6 answers, like random bimodal foams (10 of 12), and CRAX-WU-LV
is ``nacre'' in 3 of 3, like every jittered running bond (9 of 9). An open vision--language model
(Qwen2-VL-7B~\cite{wang_2024_Qwen2VL}) also sorted the images by morphological class (SI, Section~S17.5 and Fig.~S17d). An earlier single-rater round had proposed specific identities, a diatom frustule for CRAX-W$\Omega$ among them; these did not recur with three raters and random structures in the set (SI, Section~S17.6 and Fig.~S19).

A DINOv2 image embedding~\cite{oquab_2023_DINOv2} gives the same classes: random bimodal foams are seven of the seven nearest
neighbors of CRAX-W$\Omega$ and six of seven of CRAX-WU, and the nearest random structure of each main-text design is a
foam (Fig.~\ref{fig:novelty}b; SI, Section~S17.3). The distance of the three designs to the nearest canonical motif, in units of the
motifs' own spacing, is 0.78--1.74, but random structures score alike within a class, and CRAX-W$\Omega$ lies at the 35th percentile of all 240 (SI, Fig.~S17a; a CLIP embedding gives the same picture, SI, Table~S29). Distance to a motif therefore measures morphological class, foam against labyrinth,
and cannot by itself show that the optimizer converged onto the motifs or escaped from them.

The ratio of the material's characteristic length to the cell size also set the scale of the geometry
(Fig.~\ref{fig:framework}e; SI, Section~S17.2). All four designs optimized at $\ell_{ch}/L = 0.10$, for work or peak load
and over two strain windows, had 2--4 pores per cell and walls 11.5--15.2\,\% of the cell width, about $1.1$--$1.5\,\ell_{ch}$,
whereas the six optimized at 1.64 had 16--23 pores and walls of 3.6--5.9\,\%, about $0.03\,\ell_{ch}$. For the same
recipe, CRAX-MW$\Omega$ had four pores and walls 2.1 to 2.5 times thicker than CRAX-W$\Omega$ and three further runs of
its recipe.

Only the morphometrics separate the designs from random members of their class, mainly through strut
thickness and, for the two uniaxially optimized designs, anisotropy. Whether the class or the specific geometry carries the fracture
performance requires the mechanics of random foams of the same class, which remain to be run. The designs thus converge on biological motif classes without directly mimicking any one of them.

\begin{figure}[p]
  \centering
  \includegraphics[width=\linewidth]{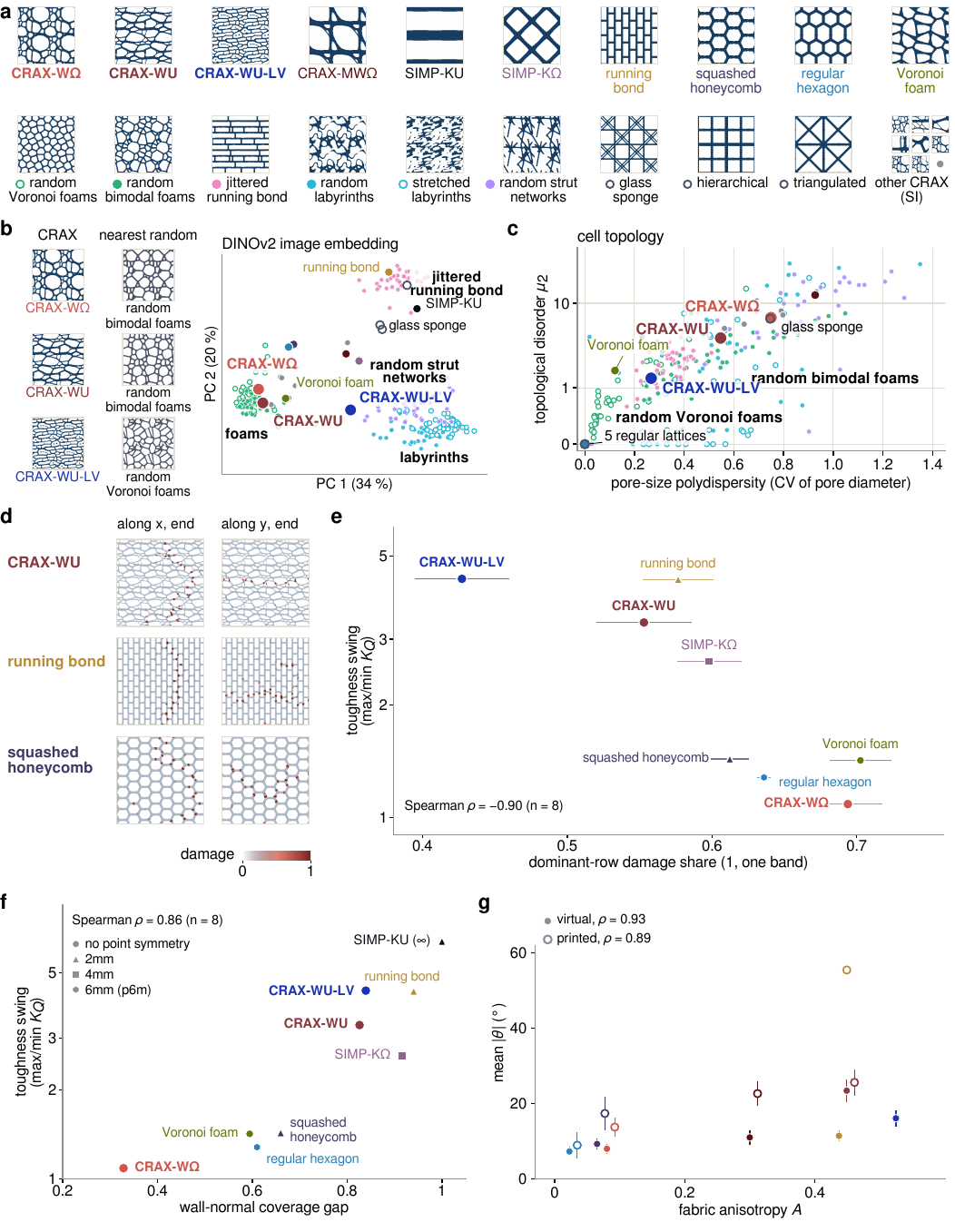}
  \caption{\textbf{Motif classes without a template, and how the designs fail.} Caption continues on the next page.}
  \label{fig:novelty}
\end{figure}

\begin{figure}[p]
  \ContinuedFloat
  \begin{minipage}[t][0.75\textheight][t]{\linewidth}
  \caption*{\textbf{Figure~\ref{fig:novelty} (continued). Motif classes without a template, and how the designs fail.}
  (a) Key to the whole figure: every cell shown in (b)--(g), as a $2\times2$ tiling above its name and symbol. Designs, SIMP baselines, and canonical cells in their colors of Fig.~\ref{fig:certification}; motifs without a design color, open rings; random families, one member each (the closest to its family's mean DINOv2 embedding), filled and open for the two members of a class; the eight other CRAX designs of the SI, one cell each, grey dots.
  (b) CRAX-W$\Omega$, CRAX-WU, and CRAX-WU-LV, each beside its nearest random structure in the DINOv2 image embedding, and all structures on the principal axes of the embedding, fitted to the 240 random structures (six families) and the seven canonical motifs.
  (c) Topological disorder $\mu_2$, the variance of the number of neighbors per pore, against pore-size polydispersity (coefficient of variation of pore diameter) for the CRAX designs, the canonical motifs, and the random structures; the glass-sponge motif matches CRAX-W$\Omega$ in these two of 21 descriptors.
  (d) Final damage of $4\times4$ periodic supercells of CRAX-WU, running bond, and the squashed honeycomb under uniaxial
  strain to 2\,\% along $x$ and along $y$ (realization 0, last archived frame at 1.9\,\%; damage dilated by one element for display).
  (e) Virtual SENT specimens (as in Fig.~\ref{fig:direction}a): toughness swing, as in (f), against the share of
  the damage in the most-damaged row of cells (1, one band; mean $\pm$ s.e.\ over 12 lattice angles, 11 for SIMP-KU and
  SIMP-K$\Omega$, and four realizations); Spearman $\rho$ over the eight designs with a transverse load path (SIMP-KU, which has
  none, is omitted); marker shape, point symmetry as in (f).
  (f) SENT toughness swing, maximum over minimum angle-mean $K_Q$ over 12 lattice angles (11 for SIMP-K$\Omega$), against the gap in the wall-normal distribution of each
  cell (0, walls in all directions); Spearman $\rho$ as in (e) (SIMP-KU, drawn at the top).
  (g) Mean crack deflection $|\theta|$ against the fabric anisotropy $A$ of the cell (0, no preferred strut direction):
  filled, virtual compact-tension specimens at $\ell_{ch}/L = 0.10$ (24 runs per design, $\pm$ s.e.); open, printed single
  cracks ($n$ = 22, 7, 6, 5, 2, and 1 for CRAX-W$\Omega$, CRAX-MW$\Omega$, the honeycomb, the hexagon, CRAX-WU, and running
  bond); $\rho$, Spearman correlation with $A$: virtual 0.93 ($p = 0.007$, seven designs), printed 0.89 ($p = 0.03$, six designs); without running bond (a single characterized specimen), printed $\rho$ is unchanged (0.90, $p = 0.08$, five designs).
  Within (d)--(f), all square cells share the same disorder realizations. Further panels: SI, Figs.~S11 (failure) and S17 (motifs); instruments and statistics: SI, Sections~S10 and S17.}
  \end{minipage}
\end{figure}

\subsection{How the designs fail}\label{sec:mechanisms}
Fig.~\ref{fig:novelty}d--g and SI, Fig.~S11 show how designs that absorb similar work can fail in very different ways (set-up: Methods; SI, Section~S10).

\textbf{Damage spread versus a single crack.} Along $x$, CRAX-WU-LV and running bond absorbed the same specific work per
cell (0.87 and $0.86\times10^{-3}$; four realizations), but by opposite routes (SI, Fig.~S11b). CRAX-WU-LV,
with 7.7 times the specific stiffness, peaked at 0.19\,\% strain and halved its load only at 3.0 times that strain.
Running bond peaked at 1.03\,\% and lost half its load on average at 1.08 times its peak strain (at most 1.12). On $4\times4$ supercells, running bond failed by one dominant crack
perpendicular to the load in each of four realizations, whereas the cracks of CRAX-W$\Omega$, CRAX-WU, and the squashed
honeycomb wandered from junction to junction and branched (Fig.~\ref{fig:novelty}d; SI, Fig.~S11a; CRAX-W$\Omega$ in
Supplementary Movies S11 and S12). In the SENT specimens,
CRAX-WU-LV spread damage most, whereas CRAX-W$\Omega$ concentrated its damage in the
most-damaged row of cells nearly as much as the Voronoi foam (SI, Table~S19) yet had the most
direction-independent toughness (\S\ref{sec:direction}). Across the eight designs with a transverse load path, the cells that spread
damage were the most direction-dependent (Spearman $\rho = -0.90$ between the dominant-row share and the toughness swing, the largest over the smallest angle-mean $K_Q$;
Fig.~\ref{fig:novelty}e): in these designs, spreading damage and crack isotropy traded off.

\textbf{A staggered motif without a template.} CRAX-WU-LV, CRAX-WU, and running bond had the lowest dominant-row shares in the
SENT specimens (SI, Table~S19; neighbors in this order lie within about one standard error). In running bond loaded
along its continuous walls, the crack stepped from wall to wall at the staggered cross-walls
(Fig.~\ref{fig:novelty}d, right column), the deflection on which staggered brick-and-mortar materials such as nacre rely~\cite{wegst_2015_Bioinspired}. The optimizer's version, which ties running bond along the optimization direction
(\S\ref{sec:validation}), was more redundant. In single-member knockouts, measured by the site robustness $r$, the peak load with the defect over the intact peak, CRAX-WU-LV had no fragile site ($r < 0.8$) and CRAX-WU five of twelve,
whereas every site of running bond was fragile (SI, Section~S10.2, Fig.~S12, and Table~S20).
Optimizing the lower tail of absorbed work thus need not cost redundancy but does not guarantee it; with walls far thinner than the material's characteristic length, where brittle solids become insensitive to flaws~\cite{gao_2003_Materials}, $r$ measures redundancy, not toughness.

\textbf{Isotropy without symmetry.} CRAX-W$\Omega$ has polydisperse closed cells without point symmetry, yet its walls
face every direction almost equally (Fig.~\ref{fig:novelty}f). The gap in its wall-normal distribution, one minus the
ratio of the thinnest to the densest $30^\circ$ sector, was 0.33, against 0.59--1.00 for the other eight designs (SI, Table~S19). Its
SENT toughness rose was the flattest of the nine designs (\S\ref{sec:direction}), as, narrowly, was its SENT work rose, and across the eight designs with a transverse load path the gap ranked the toughness swing (Spearman
$\rho = 0.86$) but not the work swing ($\rho = 0.40$; SI, Section~S10.1). Point symmetry was therefore not necessary, nor was it sufficient, since the toughness of the
square-symmetric SIMP-K$\Omega$ swung 2.62; nor was elastic isotropy, as the printed hexagon showed (\S\ref{sec:printed}). The crack
paths agree (Fig.~\ref{fig:novelty}g; SI, Fig.~S11c and Section~S10.3). In virtual compact-tension specimens the crack of CRAX-W$\Omega$ ran straight ahead at every load direction, as did those
of the hexagon and the honeycomb, consistent with the printed cracks of CRAX-W$\Omega$ (Fig.~\ref{fig:experiments}d; \S\ref{sec:printed}). In the virtual specimens, cells with directional strut fabric (CRAX-WU, CRAX-WU-LV,
running bond) deflected their cracks with load direction, by up to about $30^\circ$, and across designs the mean
deflection ranked with the anisotropy of the strut fabric (Spearman $\rho = 0.93$ virtual, 0.89 printed).

\subsection{Material properties and disorder from fracture histories}\label{sec:inverse}
The comparison with the prints rests on an assumed disorder model and a literature toughness, both of which the solver itself could calibrate. Because the solver is differentiable with respect to the material parameters as well as the design, the same gradients
can run in reverse: from fracture histories to the material that produced them, by gradient descent rather than by
Bayesian sampling~\cite{khodadadian_2020_Bayesian}. We tested this on synthetic histories at three levels of difficulty
(SI, Section~S14). With the disorder realizations known, the stiffness, toughness, and strength were recovered within 3\,\% with 5\,\% added noise (SI, Fig.~S14 and Tables~S23 and S24). With the realizations unknown, all six parameters,
including the coefficient of variation, correlation length, and strength--toughness cross-correlation of the disorder
fields, were recovered within 5\,\% from 48 histories each of a solid cell, a $2\times2$ solid supercell, and
CRAX-W$\Omega$, identified together and blind (SI, Fig.~S16 and Table~S25). The supercell was essential: without it,
the correlation length was missed by 17\,\% and the cross-correlation ran to its bound, although the fitted model still
reproduced the observations, so goodness of fit alone did not reveal the error. Specimens of different size together
thus make the disorder statistics identifiable, and printed tests of a solid, a supercell, and an architecture could calibrate the physics model used for TO.

\section{Discussion}\label{sec:discussion}
Brittle parts are seldom loaded along a single known direction and fail along their weakest one.
Within the two-dimensional, tension-driven scope of this work, CRAX-W$\Omega$, optimized over random load directions, is, to our knowledge, the first architected cell found by optimization whose stiffness, strength, absorbed work, and
apparent toughness all depend little on load direction at once (largest $D$ of 0.10 over 12 directions, against at
least 0.19 for every other tested cell), and it does so without point symmetry; earlier designs with near-isotropic fracture were prescribed rather than optimized and tested in two or three directions~\cite{wang_2025_Architected,
gao_2026_TopologyEnriched}. Printed, CRAX-W$\Omega$ matched the honeycomb and the hexagon in rose-mean specific peak
stress and work and, unlike them, showed no detectable dependence of either quantity on load direction. It also broke more cleanly than the squashed honeycomb: none of its 24 filmed prints fragmented, against four of ten
honeycombs, which fragmented only outside their weakest direction and had stored more elastic energy before failing than
five of the six intact ones (\S\ref{sec:printed}; SI, Section~S10.3). By keeping any one direction from storing surplus
energy, uniform strength appears to limit fragmentation, an emergent benefit that was not optimized for and that matters
wherever loose fragments are a hazard, as in implants or rotating machinery.

The reference cells show why this property has to be designed rather than inherited from geometry. Neither symmetry (the square-symmetric SIMP-K$\Omega$ in simulated apparent toughness; the hexagon, elastically isotropic by its six-fold symmetry, in printed work),
nor randomness (the Voronoi foam, in simulated stiffness and work), nor low direction dependence in one property (running bond, in work alone) gave low direction dependence in all
(\S\ref{sec:direction}, \S\ref{sec:printed}, and \S\ref{sec:mechanisms}), so direction-robust fracture has to be checked, or designed
for, directly. CRAX makes this tractable
because its simulator is differentiable and runs, batched, on a GPU: the fracture physics itself, sampled over disorder
realizations and load directions, supplies the gradient, without a training dataset, a surrogate model, or a template,
and the lower tail of performance, which governs the reliability of brittle parts~\cite{weibull_1951_Statistical},
becomes an explicit design target.

This uniformity was not free: along $\phi = 0$, printed CRAX-W$\Omega$ reached about half the specific peak stress and work of CRAX-WU, the cell optimized for that direction (\S\ref{sec:printed}), and across the eight designs with a transverse load path, spreading damage and crack isotropy traded off (\S\ref{sec:mechanisms}). Along a single direction, optimizing for absorbed work gave cells that matched the best canonical, bioinspired lattices
and exceeded the stiffness-optimal cells designed for the same loading, at least in the lower tail of performance, with
CRAX-WU on the work--strength Pareto front.

The optimizer also arrived at the motif classes of natural and canonical
cellular solids, closed-cell polydisperse foam and nacre-like brickwork, from a nearly uniform start and without a
template. Bioinspired design commonly transplants such motifs from an organism into a new constituent; in this study such motifs emerged
from the fracture physics and the disorder of the target material. A blind vision-LLM rater likened CRAX-W$\Omega$ to a diatom frustule, a porous shell of brittle silica that is unusually strong for its weight~\cite{hamm_2003_Architecture, aitken_2016_Microstructure}; the likeness is visual, and raters shown random structures as well placed CRAX-W$\Omega$ with closed-cell foams (\S\ref{sec:novelty}; SI, Section~S17.6). We speculate that the resemblance is not a coincidence: frustules are thought to have evolved as mechanical armor~\cite{hamm_2003_Architecture}, and in \textit{Coscinodiscus} cracks are deflected around reinforced pores~\cite{aitken_2016_Microstructure}, so their pore architecture may also have been selected to keep absorbing energy as a crack grows, which is what CRAX-W$\Omega$ was optimized for, although frustules also enable flotation, filtration, and light manipulation~\cite{musenich_2025_Revealing}.

The material's characteristic length also sets the scale of the architecture (\S\ref{sec:novelty}): designs for more
brittle cells, or larger cells of the same material, become coarser rather than self-similar. At $\ell_{ch}/L = 1.64$,
the walls are far thinner than $\ell_{ch}$, the regime in which brittle solids become insensitive to flaws~\cite{gao_2003_Materials}; in ceramic nanolattices, thin walls similarly suppress brittle fracture~\cite{meza_2014_Strong}. At $\ell_{ch}/L = 0.10$, an earlier estimate of the brittleness of the printed resin (SI, Section~S8.1), the same recipe produced CRAX-MW$\Omega$, which intriguingly resembled a skewed Kagome lattice; the skew breaks the six-fold symmetry that makes the regular Kagome, the toughest of the classical isotropic elastic--brittle lattices at low relative density~\cite{romijn_2007_Fracture, holm_2026_Mode}, elastically isotropic, consistent with the strong stiffness anisotropy of its prints. Printed, it also reached the highest mean
specific peak stress and work in Fig.~\ref{fig:experiments}f (four directions; not resolved from CRAX-W$\Omega$),
suggesting that matching the brittleness used in optimization to the material may trade isotropy for average
performance.

\phantomsection\label{sec:lim_scope}%
A few key limitations are of note: (1) Our physics model is two-dimensional (plane strain), small-strain, and quasi-static, and it describes fracture in one way, by the phase-field
cohesive-zone model with a Rankine drive under monotonic strain control. We do not consider crack closure, compressive damage,
buckling, contact, rate dependence, or fatigue, soft or ductile material models,
crushing, or impact. (2) Since no representative volume exists once softening localizes (SI, Section~S19),
the work of a periodic cell, in which periodicity enables cracks to wrap around domain edges, is a design proxy as opposed to a true homogenized
material property, and the gradients optimize that proxy: as anticipated work ratings fell 1.8- to 3.8-fold on supercells of up to
$16\times16$ cells and partly changed order, whereas strength ratings changed little (SI, Section~S8). (3) The
designs were optimized with straight-through gradient estimators at an economical discretization (20 load steps of two staggered
sweeps), because the backward pass at the converged discretization needs about 181\,GB of GPU memory (SI, Section~S7.3); re-optimizing
CRAX-W$\Omega$ with the exact adjoint changed the cell about as much as independent runs differ, but not its certified
performance (SI, Section~S13 and Fig.~S13). (4) The prints were far more brittle relative to the cell than the material used in optimization (characteristic length about 40 times
shorter, from the measured modulus and strength and a literature toughness rather than a measured one), had three cells across the gauge, and rest on two or three
specimens per design and direction, designed solid fractions, and video gauge strain for 54 of the 77 tests (SI, Section~S18.3). The
finite-dogbone simulations agree with them in rank, not in magnitude, no printed notched specimens were mechanically tested, so apparent toughness rests on virtual specimens alone, and a
test-day effect is accounted for statistically (SI, Section~S18.5 and Fig.~S23). (5) Further limitations, each with its evidence in SI Section~S19, concern the design space (single-scale periodic solid--void cells at one volume fraction, 0.35, under uniaxial strain rather than uniaxial stress or multiaxial loading); the model and the optimization (an assumed rather than calibrated disorder model, hyperparameters fixed during solver development, a moderate, non-ablated, tail-risk measure, and a square grid bias); and the evidence (run-to-run variation of about 10\,\% between seeds, with CRAX-W$\Omega$ tying the best of four; no comparison with stress-based topology optimization; and a morphology comparison that rests on AI models rather than human experts).

We envision that four extensions follow directly from our framework. (1) A coupled implicit adjoint of the converged staggered fixed
point would make gradient memory independent of the number of sweeps; together with the exact adjoint, which already needs
far less memory than the straight-through gradient (SI, Sections~S7.3
and S11.2), it would let optimization use the certification discretization and optimize cells directly as tessellated
supercells, enabling size-effect-aware designs (SI, Section~S8). Three-dimensional cells, with about 100 times more elements at our resolution,
would likely require checkpointing and memory spread across several GPUs rather
than a new method. (2) Crack closure and finite-strain kinematics would admit compressive and buckling-dominated
failure, and with them crush and impact design. Large strains in highly elastic or plastic materials would also need
adaptive remeshing, or a particle method such as the material point method, which would doubly serve to counteract mesh topology biases. (3) The optimization scope could widen to prescribed sets of loading directions and trade-offs among work, peak stress, and damage onset, as well as multi-physical functional properties. (4) The inverse identification of material properties and intrinsic disorder (see \S\ref{sec:inverse} and SI, Figs.~S14--S16) could calibrate the model on printed
tests before the design is re-optimized, starting with the toughness of the printed resin.

Furthermore, autonomous and multi-agent systems for architected and composite materials~\cite{ghafarollahi_2024_SciAgents, buehler_2026_Models}
can propose candidates faster than they can be validated. A simulator that ranks candidates under disorder, supplies
gradients, and reports tail risk would let them optimize for reliability rather than for one material realization. For this reason, we released CRAX as a library with an agent-facing interface in which forward simulation, optimization, and
certification are commands with documented inputs and outputs that an agent can call directly (Code availability).

Isotropic failure has so far been sought through symmetry, disorder, or prescribed motifs, and checked in a few
directions. Here a differentiable simulation of fracture under disorder found it by searching every direction at once,
and printed cells showed no detectable direction dependence in strength and work. By uniting fracture mechanics, reliability statistics, stochastic
optimization, and differentiable programming, CRAX makes the reliability of brittle and quasi-brittle architected
materials a design target rather than a test outcome, for applications ranging from engine components and armor to implants and nanoarchitected solids. It also offers a
roadmap: physics written in a differentiable array language such as JAX makes gradient-based topology optimization
native to the simulator, so other nonlinear, path dependent responses could be optimized the same way.

\section{Materials and Methods}\label{sec:methods}
\subsection{Design parametrization}\label{sec:param}
Element variables $x_e \in [10^{-3}, 1]$ on a $100\times100$ grid of the periodic cell are smoothed by a periodic
linear-hat filter~\cite{bourdin_2001_Filters} of radius 2.5 elements into $\tilde\rho_e$, projected
by a smoothed Heaviside function~\cite{guest_2004_Achieving}, and mapped to element properties by the power law of solid
isotropic material with penalization (SIMP)~\cite{bendsoe_1989_Optimal},
\begin{equation}\label{eq:simp}
\begin{gathered}
  \hat\rho_e = \frac{\tanh(\beta\eta) + \tanh\!\big(\beta(\tilde\rho_e - \eta)\big)}{\tanh(\beta\eta) + \tanh\!\big(\beta(1-\eta)\big)},
  \qquad \eta = 0.5,\\
  E_e = E_{\min} + (E_0 - E_{\min})\hat\rho_e^{\,p}, \quad
  G_e = \big[G_{\min} + (G_0 - G_{\min})\hat\rho_e^{\,p}\big]\,m^G_e, \quad
  f_e = \big[f_{\min} + (f_0 - f_{\min})\hat\rho_e^{\,p}\big]\,m^f_e,
\end{gathered}
\end{equation}
with $p = 3$, floors $E_{\min} = 10^{-9}E_0$, $G_{\min} = 0.05\,G_0$, and $f_{\min} = 0.02\,f_0$, disorder multipliers
$m^G_e$ and $m^f_e$ (\S\ref{sec:fields}), and the sharpness $\beta$ raised by continuation (\S\ref{sec:optimizer}; SI,
Section~S1.5). The volume fraction is fixed at 0.35 (\S\ref{sec:lim_scope}), and, unless warm-started from an earlier design, every run starts from uniform density
with a small central Gaussian dip that breaks the symmetry of the first stiffness gradient (SI, Section~S2).

\subsection{Fracture model}\label{sec:pfczm}
Fracture follows the phase-field cohesive-zone model~\cite{wu_2017_Unified} in the form of Wu and
Nguyen~\cite{wu_2018_Length} with linear softening,
\begin{equation}\label{eq:energy}
\begin{gathered}
  \Pi(u,d) = \int_\Omega \big(\omega(d) + k_{\mathrm{res}}\big)\,E\,\psi_1(\varepsilon(u))\,dV
  + \int_\Omega \frac{G_c}{\pi}\!\left(\frac{2d - d^2}{\ell} + \ell |\nabla d|^2\right) dV,\\
  \omega(d) = \frac{(1-d)^2}{(1-d)^2 + a_1 d\,(1 - d/2)}, \qquad a_1 = \frac{4\ell_{ch}}{\pi\ell}, \qquad
  \ell_{ch} = \frac{E G_c}{f_t^2},
\end{gathered}
\end{equation}
with $\psi_1$ the plane-strain elastic energy density at unit modulus and $E$, $G_c$, $f_t$, and $\ell_{ch}$ the element
values of Eq.~\ref{eq:simp}. Damage is driven by the Rankine measure $Y = \langle\sigma_1\rangle_+^2/(2E)$ of the major
principal stress of the undamaged material, without a tension--compression split, so it nucleates where the local
stress reaches the local strength. The production material has $E_0 = 210$, $\nu = 0.3$, $G_0 = 8\times10^{-4}$, and
$f_0 = 0.32$ in code units ($\ell_{ch} = 1.64$ cell widths), with $\ell = 0.025$ cell widths and
$k_{\mathrm{res}} = 10^{-9}$ (SI, Section~S1.4). The traction--separation law does not depend on $\ell$, but structural
responses do once walls are as thin as the damage band (about $\pi\ell$), as here, so every design is evaluated at the
same $\ell$ (SI, Section~S7.5; wall versus lattice-scale failure: SI, Sections~S8.1 and S9, Fig.~S10, and
Table~S18).

\subsection{Displacement-controlled staggered solve}\label{sec:forward}
Softening admits no equilibrium under load control, so each realization is driven by a monotonically increasing
displacement (finite structures) or macroscopic strain up to 2\,\% (periodic cells), in load steps of staggered sweeps:
$20\times2$ in optimization and $136\times6$ in certification. Each sweep solves for the displacement by
Jacobi-preconditioned conjugate gradients, updates the history $H \leftarrow \max(H, Y)$ at the Gauss points, and solves
for the damage ($d_{\mathrm{prev}} \le d \le 1$) by projected Newton iterations, all in double precision (SI,
Section~S1.6).

\subsection{Finite structures}\label{sec:finite}
Each of the eight standard problems (SI, Section~S7.2) was optimized with the production material at 35\,\% volume,
non-periodic disorder fields of the same statistics, and $\CVaR_{0.5}$ over batches of two realizations, that is, the
worse of the two, in four variants (uniform or stiffness-twin start; work or peak-load objective). Its stiffness-optimal
twin maximizes the pre-damage stiffness under the same boundary conditions, and both are evaluated at the same
phase-field length on a deterministic field and four disordered fields (SI, Table~S9).

\subsection{Periodic homogenization}\label{sec:homog}
In the split $u(X) = \epsbar\,X + \tilde{u}(X)$ (\S\ref{sec:canonical}), the affine part enters as a load assembled with
the damaged stiffness, so the periodic fluctuation $\tilde u$ carries zero mean strain by construction, as in computational homogenization~\cite{andreassen_2014_How}. A
load direction $\phi$ enters through the unit-trace strain $\epsbar = e\,(\cos^2\phi, \sin^2\phi, 2\sin\phi\cos\phi)$
(Voigt notation; SI, Section~S1.3 and Table~S13), and the macroscopic reaction $R = \partial\Pi/\partial e$, taken at
equilibrium and fixed damage (Hill--Mandel~\cite{hill_1963_Elastic}), is computed by automatic differentiation (SI,
Sections~S1.1 and S1.2).

\subsection{Disorder fields}\label{sec:fields}
The multipliers of Eq.~\ref{eq:simp} are unit-mean lognormal fields,
\begin{equation}\label{eq:fields}
  m^a_e = \exp\!\big(s\,z^a_e - s^2/2\big), \qquad s^2 = \ln\!\big(1 + 0.25^2\big), \qquad a \in \{f, G\},
\end{equation}
where $z^f$ and $z^G$ are standard Gaussian fields with cross-correlation 0.756 (0.75 between the multipliers) and a
shared Gaussian autocorrelation of length 0.15 cell widths. On the torus the Karhunen--Lo\`eve modes are Fourier modes,
so fields are sampled spectrally with the modes that hold 99\,\% of the field energy, a number that grows with the
domain area. Certification realizations are defined once on a $400\times400$ reference spectrum and truncated, so every
square-cell design and resolution sees the same field (SI, Section~S1.7).

\subsection{Objective and risk measure}\label{sec:objective}
Each iteration draws four fresh realizations (for CRAX-W$\Omega$, each with an independent, uniformly random load
direction) and minimizes
\begin{equation}\label{eq:cvar}
  L = -\frac{2}{k}\sum_{s \in \mathcal{K}} \frac{W_s}{W_{\mathrm{ref}}} + 0.04\cdot4\,\overline{\hat\rho(1-\hat\rho)},
  \qquad W_s = \int_0^{0.02} R_s\,de, \qquad k = 4,
\end{equation}
with $\mathcal{K}$ the $k/2$ realizations of lowest work (Eq.~\ref{eq:cvar_main}), $W_s$ integrated by the trapezoidal
rule, $W_{\mathrm{ref}}$ the work of a fixed monitor realization at the initial design, and the grey penalty averaged
over elements (SI, Section~S2). Because each iteration draws a fresh batch, the optimizer climbs the expected value of
this small-sample estimate, a tail-weighted measure between $\CVaR_{0.5}[W(\rho,\xi)]$ over the disorder distribution
and the mean. CRAX-WU-LV adds a local-volume constraint: the density averaged over every disk of radius 6 elements is at
least 0.15. Ablations of the strain window and the batch size are in SI, Section~S12 and Table~S22.

\subsection{Differentiation}\label{sec:diff}
Every linear solve is differentiated implicitly by a single adjoint solve,
\begin{equation}\label{eq:adjoint}
  K^{\top}\lambda = \bar{u}, \qquad
  \bar{\vartheta} \mathrel{+}= \lambda^{\top}\!\left(\frac{\partial f}{\partial\vartheta} - \frac{\partial K}{\partial\vartheta}\,u\right),
\end{equation}
so the conjugate-gradient iterations are never unrolled. The designs were optimized with straight-through estimators:
the forward pass is exact, the four projected-Newton iterations of each damage solve are unrolled, and the backward pass
replaces the maximum, positive-part, and clip operations by logistic weights of temperature $10^{-3}$ (SI, Sections~S1.8
and S7.4). The exact adjoint of the bound-constrained damage solve, which differentiates its stationarity conditions on
the frozen converged active set (SI, Section~S1.9), served for the gradient checks, the re-optimization, and inverse
identification. With a complementarity polish, which proved necessary for exactness, its gradients match central finite differences within 2\,\% at all five tested design states (up to 4\,\% along the adjoint gradient without the polish; SI, Fig.~S4 and Section~S7.3). The straight-through gradient pointed uphill at four of the five states
(cosine similarity 0.65--0.90 with the exact gradient) but not at the converged CRAX-W$\Omega$ loaded along $x$ (SI,
Section~S7.4 and Table~S11). Before each update, the gradient is smoothed by the same periodic linear-hat filter (a sensitivity filter). CRAX-W$\Omega$, CRAX-WU, and CRAX-WU-LV stored every load step and sweep for the
backward pass, because rematerialization (recomputing them) can change the active set of the damage update and corrupt the gradient (SI, Section~S11.2).

\subsection{Optimizer and continuation}\label{sec:optimizer}
Designs are updated by MMA~\cite{svanberg_1987_Method} with a move limit of 0.012, and a shift projection after every
step holds the volume fraction exactly. The projection sharpness starts at $\beta = 1$ and is multiplied by 1.5, up to
1024, once the design has stabilized or after at most 20 iterations (SI, Section~S2). The three CRAX designs optimized at
$\ell_{ch}/L = 1.64$ ran 1000 iterations each and CRAX-MW$\Omega$, the CRAX-W$\Omega$ recipe with
$G_0 = 5.02\times10^{-5}$ ($\ell_{ch}/L = 0.10$) and $40\times12$ load stepping, ran 600; all ended at $\beta$ of
130--195 with grey fractions of 0.20--2.15\,\% (SI, Tables~S1 and S2). Repeated runs of each recipe are compared in SI,
Section~S11.1 and Table~S21.

\subsection{Reference cells}\label{sec:canonical_methods}
The stiffness-optimal baselines are produced by the same pipeline with the pre-damage homogenized modulus as the
objective: SIMP-KU for uniaxial strain, SIMP-KU-LV with the local-volume constraint of CRAX-WU-LV, and SIMP-K$\Omega$
for the worst of four load cases. Running bond, the squashed honeycomb, and the Voronoi foam are parametric strut fields on
the square torus with every strut at least $2\ell$ thick, volume-matched in physical density except running bond which deviates slightly at 0.384; the regular hexagon requires a rectangular cell with aspect ratio $\sqrt3$ ($100\times173$ elements). Except for the designs noted below, every design, CRAX and reference
alike, is binarized by the same rule (solid where the physical density exceeds 0.01) and evaluated at $\beta = 1024$, and
comparisons are reported per unit solid volume; binarization raises the specific work of the stiffness-optimal cells by 9--10\,\%
and changes that of the CRAX cells by $-8$ to $+1$\,\%, so it does not favor CRAX (SI, Section~S3 and Table~S7). The
designs optimized at $\ell_{ch}/L = 0.10$, among them the printed CRAX-MW$\Omega$, and the warm-started and re-optimized
runs are instead binarized at physical density 0.5 (SI, Section~S17.1).

\subsection{Certification protocol and statistics}\label{sec:stats}
Designs are certified under one fixed protocol: the binarized design is projected at its native resolution,
block-replicated to $200\times200$ elements ($200\times346$ for the regular hexagon), and loaded to 2\,\% strain at
$136\times6$, which agrees with twice as many load steps within 2.4\,\% in work and 0.5\,\% in design ratios; the coarser
optimization discretization overstates work most for brittle, sharply peaked responses, so it is biased against
fracture-optimized designs (SI, Section~S5). Every design uses the floors of the optimization runs and $\ell = 0.025L$
(SI, Sections~S5 and S7.5, Fig.~S5, and Table~S12). Each design is evaluated on 48 held-out realizations under uniaxial
strain along $x$ and on 16 at each of 12 load directions ($0^\circ$--$165^\circ$ in $15^\circ$ steps), with seeds
disjoint from those of optimization. All square-cell designs see the same realizations, so comparisons are paired by
realization and, for directional data, by direction (the regular hexagon, on its own cell, is compared unpaired; SI,
Section~S6). Each design is compared on the loading it was designed for: the uniaxially optimized cells under
uniaxial strain along $x$, and CRAX-W$\Omega$ on statistics over all 12 directions. Failure mechanisms were read from
$4\times4$ periodic supercells, virtual SENT specimens of $8\times9$ cells at 12 lattice angles (11 for SIMP-KU and SIMP-K$\Omega$; SI, Fig.~S9), and
single-member knockouts (SI, Sections~S8 and S10).

For $n$ realizations with ordered works $W_{(1)} \le \dots \le W_{(n)}$, we report the specific work (per unit binary
solid volume) as a mean and as
\begin{equation}\label{eq:cvar_stat}
  \CVaR_\alpha = \frac{1}{m}\sum_{i=1}^{m} W_{(i)}, \qquad m = \lceil\alpha n\rceil, \qquad \alpha = 0.5,
\end{equation}
the negative of the usual upper-tail conditional value-at-risk of the loss $-W$, with $\alpha$ the tail fraction rather than the
confidence level. For directional data we add the worst-direction mean, the swing
$\max_\phi \overline{W}_\phi / \min_\phi \overline{W}_\phi$ of the 12 direction means, and the direction dependence $D$
(\S\ref{sec:exp_stats}) with a bootstrap interval, each for work and for peak stress, and $D$ of the homogenized Young's
modulus and of $K_Q$ (SI, Sections~S4, S9, and S18.5, and Table~S4). Differences are relative mean differences with
95\,\% paired-bootstrap percentile intervals (10,000 resamples), two-sided sign-flip randomization tests (10,000 flips)
with Holm adjustment across the references, and the number of realizations won; on the rose, whole realizations are
resampled and flipped jointly across the 12 directions (SI, Section~S6). Worst-direction comparisons carry no percentile
interval, because the percentile bootstrap of a minimum is biased; Fig.~\ref{fig:direction}c and SI, Table~S17 instead
re-select the weakest direction in each resample (SI, Table~S15). Conclusions rest on effect sizes and intervals; all
$p$-values are given in the SI (Sections~S6.1, S9--S11, and S18.5). Spearman $p$-values over eight or fewer designs are exact permutation values.

\subsection{Printed specimens}\label{sec:physical}\label{sec:exp_design}\label{sec:exp_fab}
Each design was converted to an ASTM D638-14 Type~I tensile dogbone with custom Python scripts (SI, Section~S18.1 and Table~S31). The binary cell raster,
binarized as in \S\ref{sec:canonical_methods}, was assigned a width of $13/3 = 4.333$~mm, so that three cells span the
13~mm gauge. It was tiled periodically over the specimen outline and rotated in place
by the raster angle $L$. The lattice fills the gauge and both tapers, and the 25~mm
clamped zones are solid apart from three relief slots per tab of the lattice dogbones.

We report the load direction in the cell frame, $\phi = (L - 90^\circ) \bmod 180^\circ$, so that $\phi = 0$ puts the
cell $x$ axis, the optimization direction of CRAX-WU, and $\phi = 0$ of the simulations, along the load. CRAX-W$\Omega$ was
printed in 12 directions, the squashed honeycomb in four, and the regular hexagon in three; by the mirror symmetry of the
honeycomb and the six-fold rotational and mirror symmetry of the hexagon, the latter two sets represent the full rose at
$30^\circ$ and $15^\circ$ spacing. For the hexagon this holds only approximately (SI, Section~S18.1). CRAX-WU was printed only along its optimization direction,
CRAX-MW$\Omega$ in four directions, running bond only with its bricks along the load and the solid without a direction
(SI, Table~S31). The designed solid fraction of the central 50~mm, $\bar\rho$, computed from the as-built
mask, was 0.340--0.360 for the lattices and 0.384 for running bond. $\bar\rho$ is designed, not measured: resin cured beyond the mask would raise the
printed solid fraction.

All lattice dogbones and three solid dogbones (78 specimens; three replicates per design and direction) were printed on
one build plate of a masked stereolithography printer from a pigmented photopolymer (8K Standard Resin, Space Grey;
Elegoo, Shenzhen, China; SI, Table~S31 and Fig.~S20). The vendor lists a tensile strength of
44.7\,MPa without test method or cure state, and no fracture toughness. The specimens stood on end, with the load axis along the build
direction, so that every layer interface is normal to the load. The through-thickness voids are therefore horizontal tunnels, whose
downward-facing ceilings receive light that cures through the layer above and can thicken walls that run across the load
but not walls that run along it. This cure-through was not measured. No erosion or dilation was applied to the geometry.
After printing, the specimens were washed in isopropyl alcohol, dried, and post-cured for 15~min (SI, Table~S31).
To check the effect of print orientation, four further solid dogbones were printed flat on a separate plate
with the same print settings, wash, and cure. The
three tested had a 1.32 times higher elastic modulus than the vertical solids; their lower strength (0.45 times, 95\,\% CI
0.17--1.21) is inconclusive (SI, Section~S18.5).

\subsection{Tensile tests}\label{sec:exp_test}\label{sec:exp_reduction}\label{sec:exp_records}
Specimens were loaded in tension on an Instron 68TM universal testing system (Instron, Norwood, MA, USA) with manual screw-action grips, after a
1~N preload, at a crosshead speed of 5~mm/min until the force had fallen to 5\,\% of its peak (SI, Table~S31). A 5~kN
load cell was used on the first test day and a 10~kN cell on the second. The peak load $P$ is the maximum force and the nominal peak stress is $\sigma_{\mathrm{p}} = P/A_0$, with the
gross gauge section $A_0 = 41.6$~mm$^2$. Strain was measured on the frames of the Instron Advanced Video Extensometer, which the software saved for 57 of the
77 tests (all
but the first test session): two windows of the specimen 50~mm apart in the straight 13~mm section were tracked by
normalized cross-correlation, and their separation gives
the gauge strain $\varepsilon_{\mathrm{g}}$, which excludes the tapers, fillets, grips, and machine. The elastic modulus
$E_{\mathrm{g}}$ is the Theil--Sen slope of nominal stress on $\varepsilon_{\mathrm{g}}$ between 10 and 40\,\% of $P$, and
the work per unit gauge volume $w_{\mathrm{g}}$ is the integral of nominal stress over $\varepsilon_{\mathrm{g}}$ up to the
peak, multiplied by the ratio of total to pre-peak work of the crosshead record, because a crack that crosses a window ends its track (SI,
Section~S18.3 and Table~S33; every test: SI, Fig.~S21). A track was
used if it passed four checks, of its template correlation, strain range, backward steps, and elastic bound (54 tracks). For the other 23 tests, crosshead work, displacement,
and stiffness were scaled by the median video-to-crosshead ratio of the same design or, for CRAX-WU and running bond, of the 46
lattice tracks without a grip event; one of the 23 contributes only its
peak force (SI, Section~S18.3). Specific values divide by the solid mass implied by the
design: the specific peak stress is $\sigma_{\mathrm{p}}/(\rho_{\mathrm{s}}\bar\rho)$, in kN\,m\,kg$^{-1}$, and the specific
work is $w_{\mathrm{g}}/(\rho_{\mathrm{s}}\bar\rho)$, in J\,kg$^{-1}$, with the listed resin density
$\rho_{\mathrm{s}} = 1.195$~g\,cm$^{-3}$; no ratio depends on $\rho_{\mathrm{s}}$ (SI, Section~S18.2). The specimens were not weighed. The modulus and strength of the resin come from the three vertical solid dogbones (SI, Table~S40); the modulus
from the two whose tracks passed the checks. With the fracture toughness of comparable post-cured photopolymers (that of this resin has not been
published), the measured modulus and strength (1.9\,GPa and 47.6\,MPa) give $\ell_{ch}/L \approx 0.04$ (range 0.02--0.07, up to
0.17 for a light cure; SI, Section~S8.1). Seventy-seven of the 82 printed specimens were tested (SI, Section~S18.1 and Table~S32); the five untested specimens leave at least two per
condition, and every interval uses the actual $n$. Specimens that slipped in the grips were re-gripped, retested, and
counted once. Two records show an early load drop followed by reloading; on video, the gauge section unloaded elastically
at each drop, so their gauge values are used (SI, Section~S18.3).

\subsection{Statistics of printed specimens}\label{sec:exp_stats}
For the lattice designs, the logarithm of each specific quantity was fitted by least squares with one mean per
design--direction cell and a common term for the second test day, which also absorbs the change of load cell (SI, Section~S18.4). Design means, rose means, and their ratios carry 95\,\% intervals from this model,
raw or day-adjusted. Ratios to the solid, tested on the second day only, use its own standard deviation (two degrees of
freedom). Direction means carry one-way pooled-variance $t$ intervals within each design. The elastic modulus was slightly higher on the second test day (SI, Section~S18.5), so its direction means are shown day-adjusted. Rose means weight the direction means of the squashed honeycomb and the regular hexagon by their multiplicity over the half rose. Direction dependence was tested by a one-way analysis of variance and by an upper-tail permutation test of the ratio of the largest to the smallest direction mean.  Each weakest-direction ratio is given with the two-sided $t$ test of the
model contrast, which treats both directions as chosen in advance. It is reported beside its expectation for equal,
isotropic designs, because the minimum of 12 noisy
direction means is biased low relative to that of four or three. Specific work, the optimization objective, is the primary
outcome: its two weakest-direction contrasts are Holm-adjusted, and specific peak stress is reported as secondary.
Direction dependence was summarized and compared between designs through $D$, with $D^2 = \sum_\phi w_\phi(\mu_\phi - \bar\mu_w)^2$,
where $\mu_\phi$ is the true mean log value in direction $\phi$ and the weights cover the full rose as above (equal for
the 12 directions of CRAX-W$\Omega$ and of simulated roses); $D^2$ was estimated without bias by subtracting the expected
contribution of replicate (prints) or realization (simulations) noise from the weighted variance of the observed
direction means, and $D$ is reported as its square root truncated at zero; for the deterministic Young's modulus of the homogenized cell, $D$ is the weighted standard deviation
of the log moduli. Intervals invert a studentized parametric simulation of the day model (prints) or a bootstrap over
realizations paired across directions and designs (simulations), and one-sided $p$-values for a smaller $D$ of
CRAX-W$\Omega$ are Holm-adjusted within each data set (SI, Sections~S18.4 and S18.5). Agreement with simulation is measured by Spearman rank correlation over the 14 simulated
design--directions (13 lattice ones and the solid; SI, Fig.~S24 and Table~S41). No other adjustment is used for inference.

\subsection{Implementation and compute}\label{sec:compute}
CRAX is implemented in JAX~\cite{bradbury2018jax} in double precision, since active-set thresholds of
$10^{-10}$--$10^{-12}$ and a stiffness contrast of $10^{9}$ are not resolvable in single precision. Runs checkpoint
every iteration, including the optimizer state, and resume after preemption; evaluations append one row per design and
realization and skip completed rows on restart. A production optimization run (1000 iterations) takes 30--35\,h on one A100 GPU
(NVIDIA, Santa Clara, CA, USA), and a certification evaluation about 5.5 minutes per realization on one V100S GPU (NVIDIA). The project used about
8,550 GPU-hours on the cluster through 8 October 2026 (SI, Section~S15 and Table~S26). SI, Table~S2 lists the designs in the main text.

\subsection{Use of generative AI}\label{sec:genai}
The CRAX code was written with assistance from agentic AI coding tools under the authors' direction, starting from the authors' JAX notebook prototypes. OpenAI Codex (OpenAI, San Francisco, CA, USA; GPT-5.4 and GPT-5.5; April--July 2026) generated the core solver. Several versions of Claude (Anthropic, San Francisco, CA, USA; Claude Fable~5, Claude
Fable~5.1, Claude Opus~5.5, and Claude Fable~5.2; July--November 2026), run in Claude Code, were used to write the periodic-cell formulation,
the exact adjoint of the projected damage update, and all later changes to the code; the analysis and plotting code; to prepare,
submit, and monitor simulation jobs; to search and summarize the literature; to check
the reported numbers against the stored data; and to revise and copyedit text of the manuscript and the Supplementary
Information (itemized, with the components of the core solver, in SI, Section~S15). The authors verified the code against finite differences and literature benchmarks.

Generative models were also used as measurement instruments in the morphology analysis (SI, Fig.~S17c, Sections~S17 and S17.4, and Table~S30). Blind ratings were collected from three independent instances of Claude Fable 5.1
(Anthropic), each run as a fresh Claude Code
subagent with no project context.
The blind structural descriptions placed in the joint image--text space (SI, Section~S17.6) came from one Claude Fable 5 instance under a similar blind protocol, without random
structures. Verbatim prompts, seeded image orders, the unblinding keys, and the raw answers are archived in the code repository (\texttt{provenance/} folder).
Image and text embeddings were computed with pretrained, non-generative models: DINOv2 ViT-S/14 (Meta AI, Menlo Park, CA, USA)~\cite{oquab_2023_DINOv2} and OpenCLIP ViT-B/32 (SI, Section~S17.3). The SI naming study (Section~S17.5) used the open-weight
vision--language model Qwen2-VL-7B-Instruct (Alibaba Cloud, Hangzhou, China)~\cite{wang_2024_Qwen2VL}. All models were used as
instruments; every model output except the joint image--text space (SI, Section~S17.6) is reported against random structures of the
same solid fraction, and the joint space is reported as corroboration only.

All code and text produced with AI assistance was reviewed, verified, and revised by the authors, who designed the study,
interpreted the results, and take full responsibility for the content of this article.

\section*{Data availability}
All data are archived in one Zenodo record, \url{https://doi.org/10.5281/zenodo.23674660}, under a CC BY 4.0
license: the tensile tests of the printed specimens (load records, processed video-extensometer videos with frame times,
video gauge tracks, and per-test results), photographs of the fabrication and of the broken specimens, and the simulation
outputs behind the figures and tables (load and damage histories, optimization histories, compact-tension and supercell
fields, inverse-identification runs). The final designs, run configurations, and the processed data each figure needs
are also in the code repository, \url{https://github.com/lamm-mit/crax}, with a script that downloads the Zenodo record.
This study did not generate new materials; the specimens were printed from a commercial resin (Materials and Methods), and
the specimen masks needed to reprint them are in the code repository.

\section*{Code availability}
All code is available under the Apache 2.0 license at \url{https://github.com/lamm-mit/crax}; release v1.0.4 is the version used in this work.

\section*{Author contributions}
J.A.B. and M.J.B. invented the computational strategy and devised experiments. J.A.B. wrote the code and performed computational and physical experiments. J.A.B and M.J.B wrote and revised the paper.

\section*{Competing interests}
The authors declare no competing interests.

\section*{Acknowledgments}
The authors acknowledge the MIT Office of Research Computing and Data for providing high performance computing resources
that have contributed to the research results reported within this paper. Generative AI tools were used to revise and
copyedit the text (Methods, Use of generative AI).

\section*{Funding}
J.A.B. was supported by the Department of Defense (DoD) through the National Defense Science and Engineering Graduate (NDSEG) Fellowship Program (award FA9550-23-F-0014). Part of this work was supported by the U.S. Department of Energy, Office of Science, Office of Advanced Scientific
Computing Research and Office of Basic Energy Sciences, Scientific Discovery through Advanced Computing (SciDAC)
program under the FORUM-AI project. Additional support was provided by DARPA (Scaling knowledge graphs for discovery policy and overcoming Gödelian boundaries) and the MIT Generative AI Consortium (MGAIC). 

\nocite{%
sigmund_1997_Design,winkler_2001_Experimental,fleck_2007_Damage,lu_2022_Fracture,khosravani_2023_Mixed,
brighenti_2023_Effect,hammond_2016_Stress,bao_1992_Role,qin_2009_Robustnessstrength,fernandes_2021_Mechanically,
lloyd_1982_Least,aurenhammer_1987_Power,berk_1987_Scattering,wilhelm_2003_Elasticity,zhang_1984_Fast,
hildebrand_1997_New,torquato_2002_Random,pfister_2013_Good,odgaard_1997_Threedimensional,
steinhardt_1983_Bondorientational,radford_2021_Learning,ilharco_2021_OpenCLIP,cherti_2023_Reproducible,
mantel_1967_Detection,fleiss_1971_Measuring,liang_2022_Mind,elegoo_2026_Compare,lefilament3d_2026_Elegoo,
gitman_2007_Representative,le_2010_Stressbased}

\bibliographystyle{unsrt}
\bibliography{references}

\includepdf[pages=-]{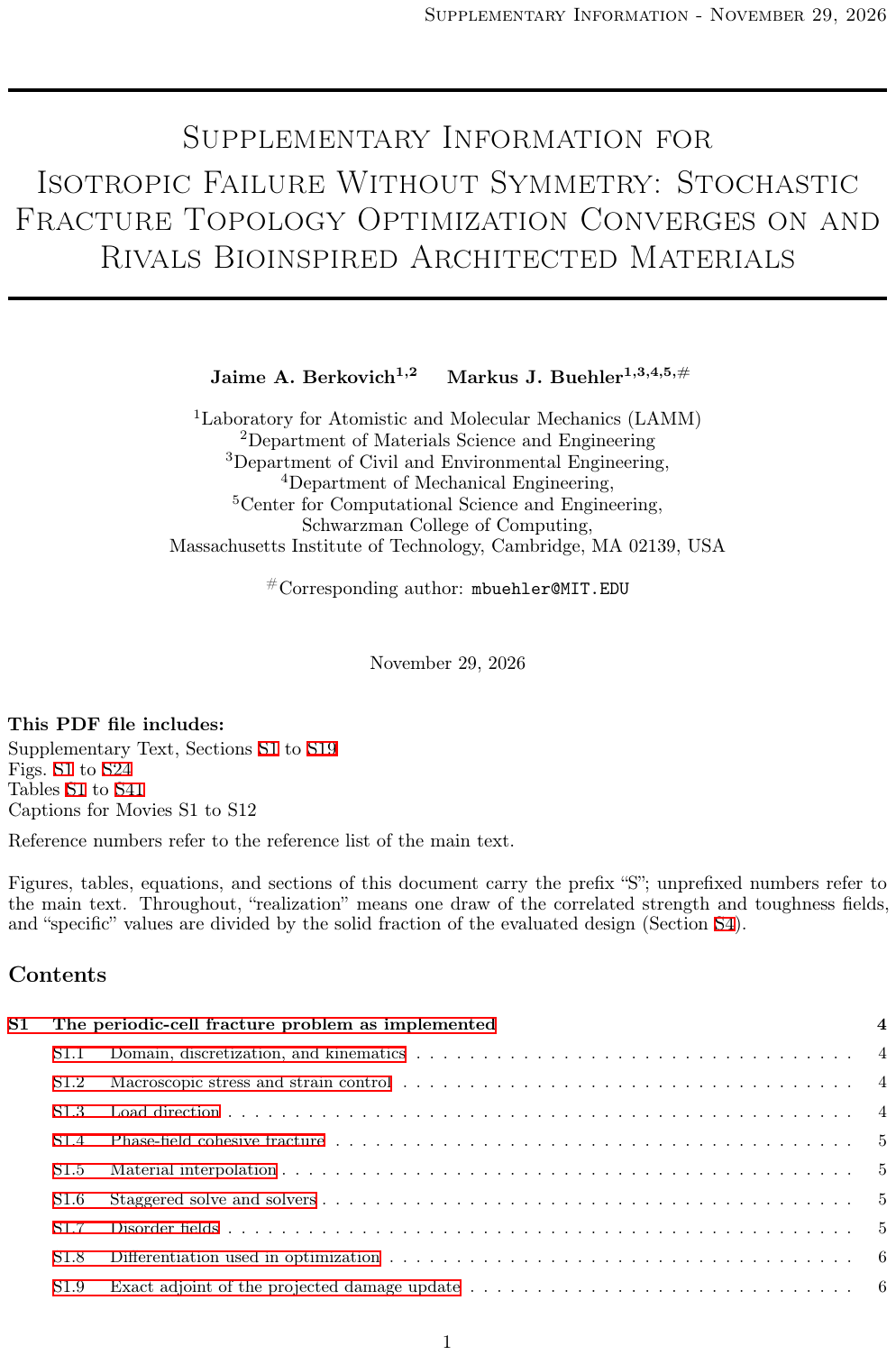}

\end{document}